\documentclass[amsmath, amssymb, aps, prl, reprint, floatfix, superscriptaddress]{revtex4-2}

\usepackage[T1]{fontenc}
\usepackage[utf8]{inputenc}

\usepackage{graphicx}
\usepackage{xcolor}
\definecolor{linkscolor}{RGB}{10,55,130}
\usepackage[pdftex, colorlinks=true, linkcolor=linkscolor, citecolor=linkscolor, urlcolor=linkscolor, breaklinks=true]{hyperref}
\usepackage{bm}
\usepackage{braket}
\usepackage{times}
\usepackage{physics}
\usepackage{units}

\usepackage{orcidlink}

\begin{document}

\title{Spontaneous Raman Scattering under Vibrational Strong Coupling: \\ The Critical Role of Polariton Spatial Mode Coherence}
\author{Maxime Dherbécourt\,\orcidlink{0009-0005-2834-4862}}
\affiliation{Université de Strasbourg, CNRS, Institut de Physique et Chimie des Matériaux de Strasbourg, UMR 7504, F-67000 Strasbourg, France}
\author{Joël Bellessa\,\orcidlink{0000-0002-9525-6898}}
\email{joel.bellessa@univ-lyon1.fr}
\author{Clémentine Symonds\,\orcidlink{0000-0003-0925-3816}}
\affiliation{Université Claude Bernard Lyon 1, CNRS, Institut Lumière Matière, UMR 5306, F-69622 Villeurbanne, France}
\author{Guillaume Weick\,\orcidlink{0000-0002-0617-5835}}
\email{guillaume.weick@ipcms.unistra.fr}
\affiliation{Université de Strasbourg, CNRS, Institut de Physique et Chimie des Matériaux de Strasbourg, UMR 7504, F-67000 Strasbourg, France}
\author{David Hagenmüller\,\orcidlink{0000-0003-4129-9312}}
\email{david.hagenmuller@ipcms.unistra.fr}
\affiliation{Université de Strasbourg, CNRS, Institut de Physique et Chimie des Matériaux de Strasbourg, UMR 7504, F-67000 Strasbourg, France}

\begin{abstract}
    {\centering(Received 10 November 2025; accepted 22 July 2026; published 8 September 2026)\par\medskip}
    \vspace{1.5mm}

    Resonant coupling of a vibration to a cavity mode has been reported to dramatically modify spontaneous Raman scattering, but subsequent studies have produced conflicting results. In this Letter, we develop a microscopic quantum framework that captures the spatial structure of polaritonic modes. In a homogeneously filled cavity, spatial overlap between polaritons and cavity resonances enforces selection rules that suppress the initially reported polaritonic Raman peaks, consistent with most experiments. In contrast, for a quasi-two-dimensional (2D) molecular layer, these rules are lifted, yielding Raman peaks at the polariton energies. Our Letter clarifies that the Raman response under vibrational strong coupling is determined by cavity-vibration spatial mode overlap and offers a framework for Raman studies of strongly coupled quasi-2D systems.

    \vspace{2mm}
    \noindent\footnotesize DOI: \href{https://doi.org/10.1103/m138-4w64}{10.1103/m138-4w64}
\end{abstract}

\maketitle

The strong light-matter coupling regime arises when the interaction strength between confined electromagnetic modes and material excitations exceeds their respective loss rates~\cite{Kimble_1998,Raimond_2001}. This regime gives rise to hybrid polariton states~\cite{hopfield_theory_1958}, which have been explored across a wide range of platforms~\cite{bellessa_strong_2004, Torma_2015,Baranov_2015,Sanvitto_Kena-Cohen_2016,Schneider2018TwoDimensional,garcia-vidal_manipulating_2021,BarraBurillo2021MicrocavityPhonon,Wright_2023}. A key feature of these hybrid states is the formation, via the optical mode, of an extended coherent superposition of material excitations. Extended polariton coherence has been directly observed~\cite{aberra_guebrou_coherent_2012} and further evidenced through energy transfer~\cite{Andrews_2000,coles_polaritonmediated_2014,zhong_energy_2017,Schafer_2019,georgiou_ultralongrange_2021, bard_extended_2022,DelPo_2021,Castagnola_2024} and polaritonic metasurfaces~\cite{chevrier_anisotropy_2019}. Strong light-matter interactions in optical cavities have attracted significant interest for their ability to influence fundamental processes such as chemical reactivity~\cite{hutchison_modifying_2012,Kowalewski_2016,Herrera_2016,flick_atoms_2017,Thomas_2019,Zhou_2021,Fregoni_2022,Mandal_2023}, transport~\cite{orgiu_conductivity_2015,Feist_2015,Schachenmayer2015,hagenmuller_2017,balasubrahmaniyam_enhanced_2023,sandik_cavityenhanced_2025}, and intermolecular interactions~\cite{haugland_intermolecular_2021,hirai_selective_2021,biswas_electronic_2025,Haugland_2025}.

In a pioneering 2015 experiment~\cite{shalabney_enhanced_2015}, it was shown that vibrational strong coupling (VSC)---the collective coupling of molecular vibrations to an optical cavity mode~\cite{Long_2015,Shalabney2015Coherent}---can profoundly alter spontaneous Raman scattering, which provides direct access to molecular vibrational excitations through inelastic light scattering processes. As a result, a typical Raman spectrum displays peaks at the vibrational mode frequencies allowed by symmetry.
By tuning a Fabry-Perot cavity into resonance with such a vibrational mode for PVAc molecules, Ref.~\cite{shalabney_enhanced_2015} reported an enhancement of the Raman signal, as compared to the bare PVAc film, of over 2 orders of magnitude and two peaks initially assigned to vibropolariton modes. This interpretation was later questioned by the same group~\cite{nagarajan_chemistry_2021} and contradicted by extensive follow-up studies using various cavity designs~\cite{takele_scouting_2021,ahn_raman_2021,menghrajani_probing_2022,verdelli_chasing_2022}, which consistently reported only a single Raman peak at the bare vibrational frequency, with no evidence of polaritonic splitting or enhanced Raman scattering.
In parallel, theoretical works based on Tavis-Cummings-like models~\cite{delpino_signatures_2015,strashko_raman_2016} with a single homogeneous cavity mode predicted polaritonic signatures in the Raman spectrum. The calculated intensities were similar to those in free space, failing to explain the large enhancement reported initially and contradicting later experiments. As a result, the mechanisms governing spontaneous Raman scattering under VSC remain highly debated and unresolved.

\begin{figure}
    \includegraphics[width=\columnwidth]{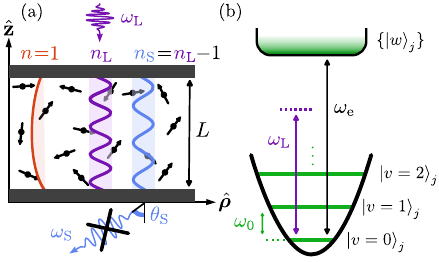}
    \caption{\label{fig:modele} 
    $N$ molecules are confined in a Fabry-Perot cavity of thickness $L$. 
    (a)~The cavity field is quantized along $\hat{\mathbf{z}}$, yielding discrete mode indices $n$. 
    Raman scattering is driven by a laser of frequency $\omega_\mathrm{L}$ and detected at frequency $\omega_\mathrm{S}$ and angle $\theta_\mathrm{S}$, corresponding to polariton modes with indices $n_\mathrm{L}$ and $n_\mathrm{S}$. 
    Selection rules suppress the resonant polariton Raman peaks with $n = 1$. 
    (b)~Each molecule $j$ has electronic and vibrational degrees of freedom: a ground-state vibrational mode $\ket{v}_j$ of frequency $\omega_0$ strongly coupled to the cavity, and excited-state vibrational modes $\ket{w}_j$ of frequencies $\omega_w$ weakly coupled to the cavity.}
\end{figure}

In this Letter, we develop a microscopic quantum framework for spontaneous Raman scattering in Fabry-Perot cavities that explicitly incorporates the spatial structure of the cavity modes [Fig.~\ref{fig:modele}(a)]. Translational invariance parallel to the cavity plane enforces conservation of in-plane momentum. Along the cavity axis, the polariton inherits the spatial profile of the cavity mode resonant with the vibrational transition. Together with the mode structure of the incident and scattered fields, the polariton profile imposes additional selection rules. For a cavity homogeneously filled with molecules as in the original experiment~\cite{shalabney_enhanced_2015}, this polaritonic spatial structure, absent in standard Tavis-Cummings-like models, suppresses resonant polaritonic Raman peaks, fully consistent with most experimental observations. 
To modify the selection rules arising from the overlap between the coherent superposition of vibrations and the cavity modes involved in Raman scattering, we extend our analysis to a quasi-2D geometry in which a thin molecular layer is embedded at a fixed position along the cavity axis. This configuration permits Raman peaks at the polariton energies, in stark contrast to the homogeneous case.
For intermediate molecular filling fractions of the cavity, we find a progressive suppression of the polaritonic Raman peaks as the filling fraction increases, eventually leading to the filled-cavity scenario where these peaks are completely absent.
In all configurations examined, we observe no enhancement of the signal compared to the nonresonant cavity case.
These results demonstrate that extended molecular coherence in polaritonic states play a pivotal role in shaping Raman scattering under VSC. Thus, our Letter makes a significant contribution toward resolving the previously reported ambiguity surrounding spontaneous Raman scattering in strongly coupled systems, initially raised in Ref.~\cite{shalabney_enhanced_2015} and examined in subsequent studies~\cite{takele_scouting_2021,ahn_raman_2021,menghrajani_probing_2022,verdelli_chasing_2022}, and lays the groundwork for applying Raman spectroscopy to investigate strongly coupled quasi-2D materials.

We consider a Fabry-Perot cavity formed by two mirrors separated by a distance $L$, which supports a set of quantized electromagnetic modes labeled by an in-plane wave vector $\mathbf{q}$ and a mode index $n \in \mathbb{N}$, which reflects the breaking of translational invariance along the cavity axis $z$ [Fig.~\ref{fig:modele}(a)]. The cavity contains $N$ identical molecules, each located at an in-plane position $\bm{\rho}_j$ and an axial position $z_j$.
We use the Born-Oppenheimer framework, where a molecular wave function is factorized into electronic and nuclear components. We restrict the electronic structure to the two lowest manifolds: a ground state with zero energy and a first excited state at energy $\hbar\omega_\mathrm{e}$. The nuclear ground-state manifold consists of a single harmonic vibrational mode with frequency $\omega_0$, while the excited-state manifold includes multiple vibrational modes $\ket{w}_j$ with energies $\hbar\omega_w$, which are nonresonantly coupled to the cavity [Fig.~\ref{fig:modele}(b)].
Since Raman scattering proceeds via virtual excited states, we focus on the ground-state vibrational mode strongly coupled to the cavity, treating excited-state vibrational modes as a weakly coupled reservoir.
Given the typically low molecular densities in experiments~\cite{shalabney_enhanced_2015,takele_scouting_2021,ahn_raman_2021,menghrajani_probing_2022,verdelli_chasing_2022}, intermolecular dipole-dipole interactions vary slowly in space and can be neglected. Furthermore, since the cavity length is much larger than the typical distance between the dipoles, the effects of image dipoles induced by the mirrors can also be safely ignored~\cite{allard_2024}.

The kinetic part of the electronic Hamiltonian for the $j$th molecule is given by $H_j^{(\mathrm{e})} = [\mathbf{P}^{(\mathrm{e})}_j + e\mathbf{A}(\bm{\rho}_j,z_j)]^2/2m_\mathrm{e}$,
where $\mathbf{P}^{(\mathrm{e})}_j$ is the electronic transition momentum, $\mathbf{A}$ is the cavity vector potential~\cite{kakazu_quantization_1994}, and $-e$ and $m_\mathrm{e}$ are, respectively, the electron charge and mass. Expanding this expression gives a kinetic term $\propto (\mathbf{P}^{(\mathrm{e})}_j)^2$, which only provides a global energy shift and is, therefore, disregarded, a light-matter interaction term $\propto \mathbf{P}^{(\mathrm{e})}_j \cdot \mathbf{A}(\bm{\rho}_j,z_j)$, and a diamagnetic term $\propto \mathbf{A}^2(\bm{\rho}_j,z_j)$. By tuning the cavity length, we ensure strong coupling to a single molecular vibrational mode, while electronic transitions that are far off resonance remain weakly coupled. This allows us to treat the electronic degrees of freedom perturbatively, while the nuclear degrees of freedom are treated exactly. Under resonant conditions, the electronic diamagnetic term is neglected, and the total Hamiltonian $H$ is partitioned into a nonperturbative part $H_0$ and a perturbation $H_1 \propto \mathbf{P}^{(\mathrm{e})}_j \cdot \mathbf{A}(\bm{\rho}_j,z_j)$.
The nonperturbative part is decomposed as $H_0 = H^\mathrm{(cav)} + H^{(\mathrm{v}0)} + H^{(\mathrm{v}1)}$, where the cavity contribution is given by $H^\mathrm{(cav)} = \sum_{\mathbf{q},n} \hbar \omega_{q,n} a_{\mathbf{q},n}^\dagger a_{\mathbf{q},n}^{\phantom{\dagger}}$. Here, $q=|\mathbf{q}|$ is the in-plane wave vector modulus, $a_{\mathbf{q},n}$ is the photon annihilation operator, and $\omega_{q,n} = c [q^2 + (\pi n/L)^2]^{1/2}$ is the cavity mode frequency, with $c$ the speed of light. The nuclear Hamiltonian is written as a sum of contributions from the excited state manifold $H^{(\mathrm{v1})} = \sum_{j=1}^N \sum_w \hbar \omega_w \ket{w}_j \bra{w}_j$, and from the ground-state vibrational mode $H^{(\mathrm{v}0)} = \sum_{j=1}^N H_j^{(\mathrm{v}0)}$, where
\begin{equation}
    H_j^{(\mathrm{v}0)} = \frac{1}{2M} \left[ \mathbf{P}_j^{(\mathrm{v}0)} - Q\mathbf{A}(\bm{\rho}_j,z_j)\right]^2 + \frac{M\omega_0^2}{2} \left(\mathbf{X}_j^{(\mathrm{v}0)}\right)^2,
\end{equation}
with, respectively, $\mathbf{X}_j^{(\mathrm{v}0)}$ and $\mathbf{P}_j^{(\mathrm{v}0)}$ the ground-state vibrational mode position and momentum operators, $M$ the reduced mass, and $Q$ the Born effective charge~\cite{Note1}.

\begin{figure*}
    \centering
    \includegraphics[width=\textwidth]{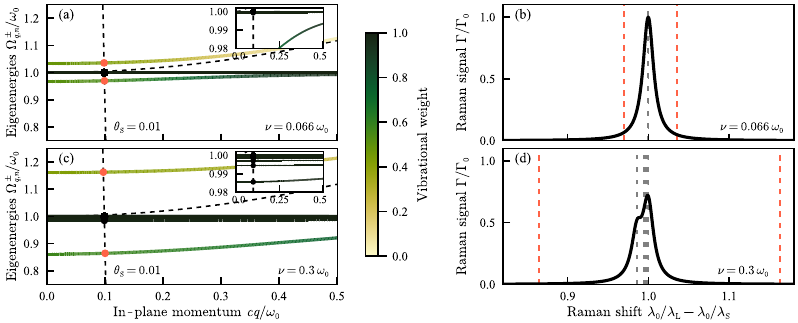}
    \caption{\label{fig:cavity_bulk}
    Absence of resonant polaritonic peaks in the Raman spectra for a Fabry-Perot cavity filled with molecules.
    Panels (a) and (c) display the polaritonic dispersion relations for two coupling strengths $\nu$, with color indicating their vibrational weight $( x^{\pm}_{q,n})^2-( z^{\pm}_{q,n})^2$ (see text).
    At fixed scattered angle $\theta_\mathrm{S}$, each dashed line corresponds to a Raman spectra in panels (b) and (d). The intersections between the dashed lines and the polaritonic branches are marked by points (red points for resonant $n=1$ polaritons, black for the ``dark'' polaritons $n>1$), which denote the predicted Raman peaks solving the energy conservation condition of Eq.~\eqref{eq:raman_rate}. These peaks are indicated by vertical dashed lines, with colors matching those of the points, in panels (b) and (d).
    Each peak is broadened by a Lorentzian of width $0.008\,\omega_0$ for clarity. The insets display enlarged regions of the dispersion.
    Peaks are normalized by the Raman signal of the off-resonant cavity $\Gamma_0$.
    Parameters: $\omega_\mathrm{L} = 10.9\, \omega_0$, $L = \lambda_0/2$, with $\lambda_0 = 2\pi c / \omega_0$ and analogous definitions for $\lambda_\mathrm{L}$ and $\lambda_\mathrm{S}$. The highest-energy cavity mode considered in the calculations is $n=100$.}
\end{figure*}

For simplicity, we model the molecules as arranged on a square lattice with their dipole moments aligned parallel to the cavity plane. To treat the strongly interacting light-matter Hamiltonian, that is $H^\mathrm{(int)} = H^{\mathrm{(cav)}} + H^{(\mathrm{v}0)}$, it is convenient to introduce the bosonic operators $b_j$, which annihilate a vibrational quantum in molecule $j$. Then, we define the collective vibrational operators
\begin{equation}\label{eq:collective_operators}
    S_{\mathbf{q},n} = \sqrt{\frac{2}{N}} \sum_{j=1}^N \mathrm{e}^{-\mathrm{i} \mathbf{q} \cdot \bm{\rho}_j} \sin\left(\frac{\pi n}{L}z_j\right) b_j \,
\end{equation}
which satisfy bosonic commutation relations in the large-$N$ limit. As the spatial profile of these collective modes along $z$ matches the cavity mode functions~\cite{Note1}, $H^\mathrm{(int)}$ can be decomposed into independent $4 \times 4$ sub-blocks, each coupling a single cavity mode to a corresponding single collective vibrational mode, such that $H^\mathrm{(int)} = \sum_{\mathbf{q},n} H^{(\mathrm{int})}_{\mathbf{q},n}$, with
\begin{align}\label{eq:H_vib_semidiag}
    H^{(\mathrm{int})}_{\mathbf{q},n} =\;& \hbar \omega_{q,n} a_{\mathbf{q},n}^\dagger a_{\mathbf{q},n}^{\phantom{\dagger}} + \hbar \omega_0 S_{\mathbf{q},n}^\dagger S_{\mathbf{q},n}^{\phantom{\dagger}} \nonumber \\
    &+ \hbar g_{q,n} \left( S_{\mathbf{q},n}^\dagger a_{\mathbf{q},n}^{\phantom{\dagger}} - S_{\mathbf{q},n} a_{-\mathbf{q},n} + \mathrm{H.c.} \right) \nonumber \\
    &+ \hbar D_{q,n} \left( a_{\mathbf{q},n}^{\phantom{\dagger}} a_{\mathbf{q},n}^\dagger - a_{\mathbf{q},n} a_{-\mathbf{q},n} + \mathrm{H.c.} \right).
\end{align}
This form of the Hamiltonian is commonly associated with a ``decoupled scenario''~\cite{chervy_room_2018,balasubrahmaniyam_coupling_2021,cortese_realspace_2023,mandal_microscopic_2023,godsi_exploring_2023,mornhinweg_sculpting_2023,tay_multimode_2025}. Such a decoupling between the different cavity modes, enabled by homogeneous molecular filling and explicit inclusion of spatial dispersion, is a key feature of our model. The coupling strengths $g_{q,n}$ and $D_{q,n}$ are characterized by the ground-state vibrational plasma frequency $\nu = (Q^2 / M \varepsilon_0 d^3)^{1/2}$, where $\varepsilon_0$ is the vacuum permittivity and $d$ is the lattice constant~\cite{Note1}. Diagonalization of the Hamiltonian~\eqref{eq:H_vib_semidiag} is achieved through a Hopfield-Bogoliubov transformation~\cite{hopfield_theory_1958}, and leads to two polariton modes ($\sigma=\pm$) $p_{\mathbf{q},n}^{\sigma} = w_{q,n}^{\sigma} a_{\mathbf{q},n} + x_{q,n}^{\sigma} S_{\mathbf{q},n} + y_{q,n}^{\sigma} a_{-\mathbf{q},n}^\dagger + z_{q,n}^{\sigma} S_{-\mathbf{q},n}^\dagger$ with frequencies $\Omega^{\sigma}_{q,n}$ in each subspace $\{\mathbf{q},\,n\}$~\cite{Note1}.

We model the spontaneous Raman scattering process by considering an incident photon of frequency $\omega_\mathrm{L}$ getting scattered into a photon of frequency $\omega_\mathrm{S}$ at an angle $\theta_\mathrm{S}$ relative to the $z$ axis, while simultaneously creating a polaritonic excitation in the system with in-plane wave vector $\mathbf{q}$ and mode index $n$. The in-plane wave vectors and mode indices of the incident and scattered photons are denoted by $\mathbf{q}_\mathrm{L}$, $n_\mathrm{L}$ and $\mathbf{q}_\mathrm{S}$, $n_\mathrm{S}$, respectively [Fig.~\ref{fig:modele}(a)].
This choice of initial and final states remains valid at room temperature, as thermal energy is typically much smaller than the vibrational energy $\hbar \omega_0$. We emphasize that, to validate our framework, we first calculated the Raman scattering rate in free space and successfully recovered the expected Raman peak at the lower polariton frequency~\cite{Note1}, in agreement with the seminal experimental results of Henry and Hopfield~\cite{henry_raman_1965}. Applying second-order perturbation theory to $H_1$ through Fermi's golden rule \cite{cohen-tannoudji_atom_1998}, we recover that in-plane momentum is conserved, $\mathbf{q} = \mathbf{q}_\mathrm{L} - \mathbf{q}_\mathrm{S}$.

\begin{figure*}
    \centering
    \includegraphics[width=\textwidth]{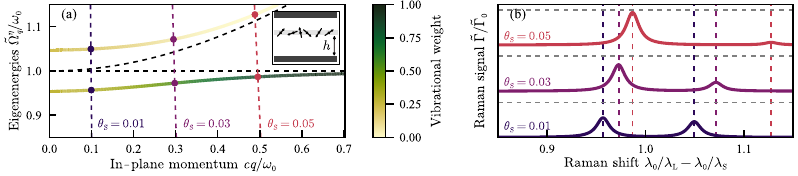}
    \caption{Raman spectra for a Fabry-Perot cavity containing a single molecular layer.
    A schematic of this configuration is shown in the inset of panel (a).
    (a) Polaritonic dispersion, with colored dashed lines indicating the Raman spectra shown in panel (b) for corresponding scattered angles $\theta_\mathrm{S}$. Intersections between dashed lines and polaritonic branches in (a) denote the predicted Raman shifts, determined by the energy conservation condition in Eq.~\eqref{eq:raman_rate_slice}, and are marked as vertical dashed lines of matching color in panel (b).
    Horizontal dashed lines in panel (b) indicate, for each $\theta_\mathrm{S}$, the Raman signal of the off-resonant cavity in the quasi-2D geometry.
    Here, we take a coupling strength of $\nu = 0.066\, \omega_0$ and fix the layer height at $h = 0.48\, L$. Other parameters are identical to Fig.~\ref{fig:cavity_bulk}. The highest-energy cavity mode considered in the calculations is $n=5$.}
    \label{fig:cavity_slice}
\end{figure*}

The resulting normalized Raman scattering rate reads
\begin{align}\label{eq:raman_rate}
    \Gamma =&\; \Gamma_0 \sum_{\sigma=\pm} \sum_n
    \delta\!\left(\left[\omega_\mathrm{L} - \omega_\mathrm{S} - \Omega^\sigma_{|\mathbf{q}_\mathrm{L} - \mathbf{q}_\mathrm{S}|,n}\right]/\omega_0\right)\nonumber\\
    &\times \left( x^\sigma_{|\mathbf{q}_\mathrm{L} - \mathbf{q}_\mathrm{S}|,n} \right)^2 f_n,
\end{align}
where the selection rule factor is \cite{Note1}
\begin{equation}\label{eq:selection_rules}
    f_n \propto \left[\int_0^L \mathrm{d}z\, \sin \left(\frac{\pi n_\mathrm{L}}{L} z\right) \sin \left(\frac{\pi n}{L} z\right)\sin \left(\frac{\pi n_\mathrm{S}}{L} z\right)\right]^2.
\end{equation}
In Eq.~\eqref{eq:raman_rate}, the Dirac delta enforces energy conservation and, thus, fixes the Raman shift. The normalization prefactor $\Gamma_0$, which depends on $\omega_\mathrm{S}$ and $\theta_\mathrm{S}$, is the Raman scattering rate of the off-resonant cavity, identical to that of an equivalent set of incoherent molecules filling the cavity. The factor $(x^\sigma_{|\mathbf{q}_\mathrm{L} - \mathbf{q}_\mathrm{S}|,n})^2$ is the vibrational weight of the polaritons, evaluated without the counter-rotating contribution~\cite{Note1}.
The integral in Eq.~\eqref{eq:selection_rules} encodes the spatial overlap between the incident, polaritonic, and scattered modes along the cavity axis. This overlap directly reflects the spatial structure of the collective vibrational operators \eqref{eq:collective_operators} and, consequently, the spatial profile of the cavity modes.
Crucially, it can be shown that Eq.~\eqref{eq:selection_rules} leads to $f_n \propto 1-\delta_{n,n_\mathrm{L}\pm n_\mathrm{S}}$, which is nonzero only for specific mode index combinations \cite{Note1}.
In particular, the resonant polaritonic branches ($n=1$) are always suppressed.
These selection rules are a core contribution of our model, as it directly reflects the influence of cavity geometry and the inclusion of the polariton spatial coherence.

Figure~\ref{fig:cavity_bulk} presents the dispersion relations of the coupled cavity-vibrational system described by Eq.~\eqref{eq:H_vib_semidiag} for every mode index $n$, along with their corresponding Raman spectra for two different coupling strengths $\nu$. To facilitate comparison with experimental data, we adopt parameters similar to those in Ref.~\cite{ahn_raman_2021}, specifically tuning the cavity length $L$ to ensure resonance between the vibrational energy and the first cavity mode $n=1$ at zero in-plane wave vector, i.e., $\omega_0 \simeq \omega_{\mathbf{q}=\mathbf{0},n=1}$.
The dispersion relations in Figs.~\ref{fig:cavity_bulk}(a) and \ref{fig:cavity_bulk}(c) reveal two resonant lower (LP) and upper (UP) polaritonic branches originating from the coupling to the cavity mode $n=1$, along with additional ``dark'' branches, predominantly vibrational in character, that arise from the off-resonant coupling to higher-order cavity modes with $n>1$.

The predicted Raman shift of the peaks at fixed scattered angle $\theta_\mathrm{S}$, indicated by intersections along the dashed line in the dispersion plots, are marked as dashed lines in the Raman spectra of Figs.~\ref{fig:cavity_bulk}(b) and \ref{fig:cavity_bulk}(d). For both coupling strengths, a prominent central feature near the vibrational frequency arises, composed of the superposition of multiple peaks associated with the dark branches.
The height of this bare vibrational peak is the same as that of the off-resonant cavity.
Crucially, the two resonant polariton modes ($n=1$), although present in the dispersion and, therefore, detectable by infrared spectroscopy, do not give rise to observable Raman peaks. This absence arises from the selection rules encoded in Eq.~\eqref{eq:selection_rules}, which, notably, are independent of the coupling strength.
This result highlights the critical role of cavity geometry and polaritonic structure in determining the observable Raman response, which were disregarded in previous theoretical studies~\cite{delpino_signatures_2015,strashko_raman_2016}. Finally, we note the presence of a shoulderlike feature in Fig.~\ref{fig:cavity_bulk}(d) corresponding to the dark branch $n=2$, which becomes more pronounced at higher coupling strengths and may account for similar observations reported in previous experiments~\cite{ahn_raman_2021}.

The suppression of the two resonant polariton Raman peaks can be straightforwardly understood by considering the case of a normal scattering angle $\theta_\mathrm{S} = 0$, where the in-plane momentum $\mathbf{q} = \mathbf{q}_\mathrm{L} - \mathbf{q}_\mathrm{S}$ vanishes, allowing for analytical treatment. Energy conservation encapsulated in Eq.~\eqref{eq:raman_rate} yields a Raman shift for the LP and UP resonant polaritons deviating from the vibrational frequency by $\pm\, \nu/2$.
Expressing this in terms of mode indices gives that $n_\mathrm{L} - n_\mathrm{S} = 1 \pm \nu/2\omega_0$. Thus, for $\nu/2\omega_0$ not too large, we find that $n_\mathrm{L} - n_\mathrm{S} \simeq 1$, leading to the suppression of the two resonant $n=1$ Raman peaks as encoded in the selection rules \eqref{eq:selection_rules}. This result can be generalized to any cavity mode index resonant with the vibrational frequency, showing that the corresponding resonant Raman peaks are always suppressed~\cite{Note1}.

Now, we turn to the case of a cavity containing a single molecular layer parallel to the cavity mirrors and fixed at a height $z_j = h$. In contrast to the previously considered filled cavity, the $z_j$ dependence of the vector potential $\mathbf{A}(\bm{\rho}_j,z_j)$ is now constant for all molecules.
As a result, the collective vibrational operators can no longer carry an $n$ dependence if bosonic commutation relations are to be preserved \cite{balasubrahmaniyam_coupling_2021,mandal_microscopic_2023,tay_multimode_2025,koshkaki_exciton_2025}.
Instead, they reduce to purely in-plane collective operators, defined as $\tilde{S}_{\mathbf{q}} = \frac{1}{\sqrt{N}} \sum_{j=1}^N \mathrm{e}^{-\mathrm{i} \mathbf{q} \cdot \bm{\rho}_j} b_j$.
In contrast to the ``decoupled'' Hamiltonian~\eqref{eq:H_vib_semidiag} for the filled cavity, the absence of $n$ dependence in these collective operators allows each vibrational mode to couple in principle to all cavity modes $n$ (yet predominantly to the energy-resonant one). Thus, the Hamiltonian $H^\mathrm{(int)}$ can no longer be diagonalized independently within each subspace $\{\mathbf{q},n\}$. 

Numerical diagonalization via a Hopfield-Bogoliubov transformation~\cite{hopfield_theory_1958} introduces polariton operators $\tilde{p}_{\mathbf{q}}^{\eta} = \sum_n \tilde{w}_{q,n}^\eta a_{\mathbf{q},n} + \tilde{x}_{q}^\eta \tilde{S}_{\mathbf{q}} + \sum_n \tilde{y}_{q,n}^\eta a_{-\mathbf{q},n}^\dagger + \tilde{z}_{q}^\eta \tilde{S}_{-\mathbf{q}}^\dagger$ with frequencies $\tilde{\Omega}_{q}^{\eta}$, labeled by the index $\eta$~\cite{Note1}. Figure~\ref{fig:cavity_slice}(a) shows the corresponding polariton dispersion near the bare vibrational energy. Importantly, the coupling between the different cavity modes $n$ eliminates dark branches from the polariton spectrum close to the bare vibrational mode frequency. As before, the mapping between dispersion and Raman spectra at various scattered angles $\theta_\mathrm{S}$ is indicated by dashed lines, which provide the expected Raman shifts for each polariton.

The spontaneous Raman scattering rate for the single-layer configuration is obtained analogously to the filled cavity case~\cite{Note1}, with in-plane momentum conservation and axial momentum unconstrained. The normalized rate reads 
\begin{eqnarray}\label{eq:raman_rate_slice}
    \tilde{\Gamma} &=& \tilde{\Gamma}_0\, \sum_\eta
    \delta\!\left(\left[\omega_\mathrm{L} - \omega_\mathrm{S} - \tilde{\Omega}_{|\mathbf{q}_\mathrm{L} - \mathbf{q}_\mathrm{S}|}^{\eta} \right]/\omega_{0}\right)  \left( \tilde{x}_{|\mathbf{q}_\mathrm{L} - \mathbf{q}_\mathrm{S}|}^{\eta} \right)^2
\end{eqnarray}
which closely resembles Eq.~\eqref{eq:raman_rate} but lacks the selection rule factor $f_n$ due to the absence of spatial structure in the collective vibrational operators. The normalization prefactor $\tilde{\Gamma}_0$ is the Raman scattering rate of the off-resonant cavity in the single-layer configuration, which differs from $\Gamma_0$ due to the different cavity geometry and mode structure \cite{Note1}.
The corresponding Raman spectra, shown in Fig.~\ref{fig:cavity_slice}(b), reveal that both resonant polaritonic peaks are now observable, unlike in the filled cavity scenario.
As for the peak heights, each peak reaches $1/2$ when $\theta_\mathrm{S} \ll 1$ and the total never exceeds $1$, relative to the nonresonant (incoherent) case, thus implying no signal enhancement in such a cavity.
This again highlights the strong influence of the overlap between the cavity geometry and the collective vibrational modes on the Raman response. 
This result is particularly noteworthy when compared to Raman spectroscopy of polar crystals with coherence lengths comparable to the sample size (which is not the case for molecular ensembles) in free space, where typically only the LP can be observed experimentally due to energy and momentum conservation~\cite{henry_raman_1965}.

To examine the transition from filled-cavity to single-layer configurations, we investigate intermediate filling fractions in Ref.~\cite{Note1}, following studies on spatial mode structure effects~\cite{kunyang_exploring_2025,koshkaki_exciton_2025,mandal_microscopic_2023}.
Starting from the single-layer configuration, the Raman peaks at the polariton frequencies are progressively suppressed as the filling fraction increases, until the filled-cavity scenario is reached, where these peaks are entirely absent.
Therefore, observing a sizable Raman signal at the polariton frequencies requires optimizing both the layer height, thickness and the laser frequency~\cite{Note1}.

In conclusion, we have developed a microscopic model capable of describing spontaneous Raman scattering in Fabry-Perot cavities, which explicitly incorporates the spatial structure of the cavity modes. For a homogeneously filled cavity, our analysis reveals that selection rules imposed by the mode structure suppress the resonant polaritonic Raman peaks, in contrast to the observations of Ref.~\cite{shalabney_enhanced_2015}, but in agreement with subsequent experimental observations~\cite{takele_scouting_2021,ahn_raman_2021,menghrajani_probing_2022,verdelli_chasing_2022}. In contrast, for a single molecular layer, the polaritonic Raman peaks become accessible, highlighting the critical role of cavity geometry and polariton mode structure, which were absent in previous theoretical models.
These results provide a unified framework for interpreting multiple experimental outcomes and suggest possible future experimental configurations to probe and control the polaritonic Raman peaks in strongly coupled quasi-2D materials or thin molecular layers, with a potential enhancement of the Raman signal.
Future directions of our Letter include the study of nonlinear effects~\cite{wang_large_2021} such as stimulated Raman scattering and alternative cavity geometries such as plasmonic cavities with evanescent mode profiles~\cite{menghrajani_probing_2022}.

\begin{acknowledgments}
    \textit{Acknowledgments}---We acknowledge helpful discussions with Elo\"ise Devaux and Thomas Ebbesen. This work of the Interdisciplinary Thematic Institute QMat, as part of the ITI 2021-2028 program of the University of Strasbourg, CNRS, and Inserm, was supported by IdEx Unistra (Project No.\ ANR 10 IDEX 0002), and by SFRI STRAT'US Projects No.\ ANR-20-SFRI-0012 and No.\ ANR-17-EURE-0024 under the framework of the French Investments for the Future Program.
\end{acknowledgments}

\begin{acknowledgments}
    \textit{Data availability}---The data that support the findings of this article are not publicly available. The data are available from the authors upon reasonable request.
\end{acknowledgments}

\footnotetext[1]{See Supplemental Material for the exact derivation and diagonalization of the Hamiltonian, the calculation and numerical evaluation of the Raman scattering rate, in the case of the filled cavity, single layer, intermediate cavity filling, and free space configurations.}

\bibliography{biblio}

\end{document}


\title{Supplemental Material for\\ Spontaneous Raman Scattering under Vibrational Strong Coupling: \\ The Critical Role of Polariton Spatial Mode Coherence}

\author{Maxime Dherbécourt}
\affiliation{Université de Strasbourg, CNRS, Institut de Physique et Chimie des Matériaux de Strasbourg, UMR 7504, F-67000 Strasbourg, France}
\author{Joël Bellessa}
\email{joel.bellessa@univ-lyon1.fr}
\author{Clémentine Symonds}
\affiliation{Université Claude Bernard Lyon 1, CNRS, Institut Lumière Matière, UMR 5306, F-69622 Villeurbanne, France}
\author{Guillaume Weick}
\email{guillaume.weick@ipcms.unistra.fr}
\affiliation{Université de Strasbourg, CNRS, Institut de Physique et Chimie des Matériaux de Strasbourg, UMR 7504, F-67000 Strasbourg, France}
\author{David Hagenmüller}
\email{david.hagenmuller@ipcms.unistra.fr}
\affiliation{Université de Strasbourg, CNRS, Institut de Physique et Chimie des Matériaux de Strasbourg, UMR 7504, F-67000 Strasbourg, France}


\begin{abstract}
In this Supplemental Material, we provide details on the theoretical framework and numerical methods presented in the main text. We begin in Sec.~\ref{sec:bulk_cavity} by discussing the model of a cavity filled with molecules, including the diagonalization of the unperturbed Hamiltonian and the analytical calculation of the Raman scattering rate using perturbation theory.
Then in Sec.~\ref{sec:single_layer} we consider a single molecular layer in a cavity and numerically compute the corresponding Raman spectra.
In Sec.~\ref{sec:intermediate_filling} we investigate the intermediate situation of a cavity filled with a finite number of molecular layers and analyze the transition from the single layer (discussed in Sec.~\ref{sec:single_layer}) to the bulk case (considered in Sec.~\ref{sec:bulk_cavity}).
Finally, in Sec.~\ref{sec:fs} we apply the same formalism to a spatially coherent molecular ensemble in free space and confront our numerical results with existing experiments in order to benchmark our model.
\end{abstract}

\maketitle

\tableofcontents

\section{Cavity filled with molecules}\label{sec:bulk_cavity}
\subsection{Model}
We consider a Fabry-Perot cavity of volume $V$ formed by two perfect mirrors parallel to the $xy$ plane and separated by a distance $L$.  
The cavity is filled with $N$ identical molecules in vacuum. 
Within the Born-Oppenheimer approximation, we model each molecule by separating its electronic and nuclear degrees of freedom. A schematic representation of the setup is shown in Fig.~1(a) of the main text. The Hamiltonian of the system can be written as 
\begin{equation}
    \label{eq:H_full}
    H= H^{(\mathrm{cav})}+\sum_{j=1}^N H_{j}^{(\mathrm{e})}+\sum_{j=1}^N H_{j}^{(\mathrm{nuc})}. 
\end{equation}
Here, $H^{(\mathrm{cav})}$ corresponds to the Hamiltonian of the cavity photons, while $\sum_{j=1}^N H_{j}^{(\mathrm{e})}$ and $\sum_{j=1}^N H_{j}^{(\mathrm{nuc})}$ describe the electronic and nuclear degrees of freedom of the molecules coupled to the cavity electromagnetic field, respectively.

\subsubsection{Cavity photons}\label{subsec:bulk_cavity_h}
    In the configuration displayed on Fig.~1(a) of the main text, the photonic cavity modes are quantized in the $z$ direction. We characterize these modes by a wavevector $\mathbf{q}=\mathbf{q}_\parallel+ q_z\hat{\mathbf{z}}$ that can be decomposed into an in-plane $\mathbf{q}_\parallel = q_x \hat{\mathbf{x}} + q_y \hat{\mathbf{y}}$ and an axial component $q_z = n\pi/L$, with $n \in\mathbb{N}$. The Hamiltonian of the cavity reads
    \begin{equation}
        \label{eq:H_cav_1st}
        H^{(\mathrm{cav})} = \sum_{\mathbf{q}_\parallel, n} \sum_{\nu=1,\,2} \hbar \omega_{q_\parallel,n} a_{\mathbf{q}_\parallel,n}^{\nu\dagger} a_{\mathbf{q}_\parallel,n}^{\nu},
    \end{equation}
    where $a_{\mathbf{q}_\parallel,n}^{\nu\dagger}$ ($a_{\mathbf{q}_\parallel,n}^{\nu}$) are the creation (annihilation) operators for the cavity mode with in-plane wavevector $\mathbf{q}_\parallel$, mode number $n$, and polarization $\nu$. The corresponding dispersion relation reads 
    \begin{equation}
    \omega_{q_\parallel,n} = c\, \sqrt{q_\parallel^2 + \left(\frac{n\pi}{L}\right)^2},
    \end{equation}
    with $c$ the speed of light in vacuum.

    Within the Coulomb gauge \cite{cohen-tannoudji_atom_1998}, the transverse vector potential $\mathbf{A}(\bm{\rho},z)$ evaluated at an in-plane $\bm{\rho} = x \hat{\mathbf{x}} + y \hat{\mathbf{y}}$ and axial $z$ position can be expressed in terms of the mode spatial functions $\mathbf{u}^\nu_{\mathbf{q}_\parallel, n}(\bm{\rho},z)$ as
    \begin{equation}
        \mathbf{A}(\bm{\rho},z) = \sum_{\mathbf{q}_\parallel, n} \sum_{\nu=1,\,2} \sqrt{\frac{\hbar}{2\varepsilon_0 \omega_{q_\parallel,n} V}} \left[\mathbf{u}^{\nu}_{\mathbf{q}_\parallel, n}(\bm{\rho},z)\, a_{\mathbf{q}_\parallel, n}^{\nu} + \mathbf{u}^{\nu *}_{\mathbf{q}_\parallel, n}(\bm{\rho},z)\, a_{\mathbf{q}_\parallel, n}^{\nu\dagger}\right], 
    \end{equation}
    where $\varepsilon_0$ is the vacuum permittivity.
    Solving Maxwell's equations in our Fabry-Perot cavity using vanishing boundary conditions at the mirrors leads to the mode spatial functions  \cite{kakazu_quantization_1994}
    \begin{subequations}
        \label{eq:mode_functions}
        \begin{align}
            \mathbf{u}_{\mathbf{q}_\parallel, n}^1(\bm{\rho},z) &= \mathrm{e}^{\mathrm{i} \mathbf{q}_\parallel \cdot \bm{\rho}} \left[\mathrm{i} \sqrt{2} \sin \left(\frac{n\pi}{L}z \right) \cos \theta_{\mathbf{q}_\parallel,n} \left(\cos \varphi_{\mathbf{q}_\parallel,n} \hat{\mathbf{x}} + \sin \varphi_{\mathbf{q}_\parallel,n} \hat{\mathbf{y}} \right) - \sqrt{\frac{2}{1+\delta_{n,0}}} \cos \left(\frac{n\pi}{L}z \right) \sin \theta_{\mathbf{q}_\parallel,n} \hat{\mathbf{z}} \right],  \\
            \mathbf{u}_{\mathbf{q}_\parallel, n}^2(\bm{\rho},z) &= \mathrm{i} \sqrt{2}\, \mathrm{e}^{\mathrm{i} \mathbf{q}_\parallel \cdot \bm{\rho}} \sin \left(\frac{n\pi}{L}z \right) \left( \cos \varphi_{\mathbf{q}_\parallel,n} \hat{\mathbf{y}} - \sin \varphi_{\mathbf{q}_\parallel,n} \hat{\mathbf{x}} \right),
        \end{align}
    \end{subequations}
    with the cavity wavevector given in spherical coordinates as 
    $\mathbf{q} = q\, (\sin\theta_{\mathbf{q}} \cos\varphi_{\mathbf{q}} \hat{\mathbf{x}} + \sin\theta_{\mathbf{q}} \sin\varphi_{\mathbf{q}} \hat{\mathbf{y}} + \cos\theta_{\mathbf{q}} \hat{\mathbf{z}})$.

    In what follows, we shall assume for simplicity that the vibrational and electronic dipoles of every molecule are aligned in the $x$ direction. 
    Moreover, we choose $\varphi_{\mathbf{q}} = 0$ in Eq.~\eqref{eq:mode_functions} so that only one photon polarization ($\nu = 1$) is relevant in the sequel.\footnote{We have checked that such a simplification does not change the main result of our paper, that is the selection rule (5) of the main text.} We thus drop the $\nu$ index in the following.
    We note that the $n=0$ mode can be ignored as it does not couple to the vibrational states and thus does not contribute to the Raman scattering process.
    The only relevant part of the transverse vector potential hence reads, after identifying $\cos \theta_{\mathbf{q}_\parallel,n}$ with $\omega_{0,n} / \omega_{q_\parallel,n}$, as
    \begin{equation}
        \mathbf{A}(\bm{\rho},z) \cdot \hat{\mathbf{x}} = \mathrm{i}\sum_{\mathbf{q}_\parallel,n} \sqrt{\frac{\hbar}{\varepsilon_0 \omega_{q_\parallel,n} V}} \frac{\omega_{0,n}}{\omega_{q_\parallel,n}} \left(\mathrm{e}^{\mathrm{i} \mathbf{q}_\parallel \cdot \bm{\rho}}\, a_{\mathbf{q}_\parallel, n} - \mathrm{e}^{-\mathrm{i} \mathbf{q}_\parallel \cdot \bm{\rho}} \, a_{\mathbf{q}_\parallel, n}^{\dagger}\right) \sin \left(\frac{n\pi}{L}z \right).
    \end{equation}
    Within the above hypothesis, the cavity Hamiltonian \eqref{eq:H_cav_1st} reduces to
    \begin{equation}
    \label{eq:H_cav_simple}
        H^{(\mathrm{cav})} = \sum_{\mathbf{q}_\parallel,n} \hbar \omega_{q_\parallel,n} a_{\mathbf{q}_\parallel,n}^{\dagger} a_{\mathbf{q}_\parallel,n}^{\phantom{\dagger}}.
    \end{equation}

\subsubsection{Description of the molecules and their coupling to the cavity photons}\label{subsec:molecules_desc}
    We characterize the position $\mathbf{r}_j$ of each molecule $j=1,\ldots,N$ by an in-plane $\bm{\rho}_j = x_j \hat{\mathbf{x}} + y_j \hat{\mathbf{y}}$ and axial $z_j \hat{\mathbf{z}}$ component. 
    For simplicity, we assume that the molecules are arranged on a cubic lattice with lattice constant $d$.
    We denote by $N_\parallel$ the number of molecules in the $xy$ plane and by $N_z$ the number of molecules along the $z$ direction, so that $N=N_\parallel N_z$. In this work, we always consider the thermodynamic limit of a large number of molecules $N \gg 1$. 
    The kinetic part of the electronic Hamiltonian of molecule $j$ is given by the minimal coupling Hamiltonian in the Coulomb gauge under the dipolar approximation
    \begin{equation}
        \label{eq:H_e_j_1st}
        H_j^{(\mathrm{e})} = \frac{1}{2m_\mathrm{e}}\left[P^{(\mathrm{e})}_j\hat{\mathbf{x}} +e\mathbf{A}(\mathbf{r}_j)\right]^2,
    \end{equation}
    where $P^{(\mathrm{e})}_j$ is the momentum associated to the electronic transitions, while $-e$ and $m_\mathrm{e}$ are respectively the electron charge and mass. We restrict ourselves to the two first electronic manifolds of each molecule, that is the ground state with zero energy and the first excited state with energy $\hbar \omega_\mathrm{e}$. As we tune the cavity height $L$ such that it weakly couples to the electronic degrees of freedom, we neglect the diamagnetic term $\propto \mathbf{A}^2$ in Eq.~\eqref{eq:H_e_j_1st}. Moreover, we disregard the kinetic term $\propto(P^{(\mathrm{e})})^2$ as it only contributes a global energy shift.

    The nuclear states of the molecules in their electronic ground state are modeled as a single vibrational mode of the ionic charges along the $x$ direction. Such a vibration creates a dipole with an effective mass $M$ and Born effective charge $Q$ oscillating at the frequency $\omega_0$. We label these states as $\ket{v}_j$. As the first excited state is composed of many vibrational modes nonresonantly coupled to the cavity, we describe them by a set of modes $\ket{w}_j$ with energies $\hbar \omega_w$. We do not consider any interaction between these set of states in the two electronic manifolds. The nuclear Hamiltonian of the molecule $j$ then reads
    \begin{equation}
        \label{eq:H_v1}
        H_j^{(\mathrm{nuc})} = H_j^{(\mathrm{v}0)} + H_j^{(\mathrm{v}1)},
    \end{equation}
    with the excited state nuclear part
    \begin{equation}
        \label{eq:H_j_v1}
        H_j^{(\mathrm{v}1)} = \sum_w \hbar \omega_w \ket{w}_j \bra{w}_j,
    \end{equation}
    and the ground-state vibrational part, given in Eq.~(1) of the main text, that reads
    \begin{equation}
        \label{eq:H_v0}
        H_j^{(\mathrm{v}0)} =\frac{1}{2M} \left[ P_j^{(\mathrm{v}0)}\hat{\mathbf{x}} - Q\mathbf{A}(\mathbf{r}_j)\right]^2 + \frac{M\omega_0^2}{2} \left(X_j^{(\mathrm{v}0)}\right)^2,
    \end{equation}
    where we defined the position $X_j^{(\mathrm{v}0)}$ and momentum $P_j^{(\mathrm{v}0)}$ operators associated with the ground-state vibrational component. 
    We next introduce the ladder operators $b_j$ ($b_j^\dagger$) that annihilate (create) a vibrational excitation in the ground state manifold of molecule $j$,
    \begin{subequations}
        \begin{align}
            b_j &= \sqrt{\frac{M\omega_0}{2\hbar}} \left(X_j^{(\mathrm{v}0)} + \frac{\mathrm{i}}{M\omega_0} P_j^{(\mathrm{v}0)}\right),\\  b_j^\dagger &= \sqrt{\frac{M\omega_0}{2\hbar}} \left(X_j^{(\mathrm{v}0)} - \frac{\mathrm{i}}{M\omega_0} P_j^{(\mathrm{v}0)}\right),
        \end{align}
    \end{subequations}
    that follow the standard bosonic commutation relations $[b_j,b_{j'}^\dagger] = \delta_{jj'}$ and $[b_j,b_{j'}] = [b_j^\dagger,b_{j'}^\dagger] = 0$. The position and momentum operators then become
    \begin{subequations}
        \begin{align}
        X_j^{(\mathrm{v}0)} &= \sqrt{\frac{\hbar}{2M\omega_0}} \left(b_j^{\dagger} + b_j\right),\\
        P_j^{(\mathrm{v}0)} &= \mathrm{i}\, \sqrt{\frac{\hbar M \omega_0}{2}} \left(b_j^{\dagger} - b_j\right).
        \end{align}
    \end{subequations}

\subsubsection{Full Hamiltonian}\label{subsec:bulk_cavity_full_h}
    Summing explicitly every terms, the total Hamiltonian \eqref{eq:H_full} of the full system can be decomposed, up to a renormalization of the energies, as 
    $H=H_0+H_1$, with ($H_0$) $H_1$ a (non)perturbative part. 
    The nonperturbative part is $H_0 = H^\mathrm{{(int)}}_0 + H^{(\mathrm{v}1)}_0$ with $H^\mathrm{{(int)}}_0 = H_0^\mathrm{(cav)} + H_0^{(\mathrm{v}0)}$ [see Eqs.~\eqref{eq:H_cav_simple}, \eqref{eq:H_v1}, and \eqref{eq:H_v0}]. 
    The latter Hamiltonian can be expressed as 
    \begin{align}
        H^\mathrm{{(int)}}_0 =\;& \sum_{\mathbf{q}_{\parallel}} \sum_{n=1}^{n_\mathrm{c}} \hbar \omega_{q_{\parallel},n} a_{\mathbf{q}_{\parallel},n}^\dagger a_{\mathbf{q}_{\parallel},n}^{\phantom{\dagger}}
        + \sum_{j=1}^N \hbar \omega_0 b_j^\dagger b_j^{\phantom{\dagger}} \nonumber\\
        &+ \sqrt{\frac{2}{N}}\sum_{j=1}^N \sum_{\mathbf{q}_{\parallel}} \sum_{n=1}^{n_\mathrm{c}} \hbar g_{q_{\parallel},n}  \left[\mathrm{e}^{\mathrm{i} \mathbf{q}_{\parallel} \cdot \bm{\rho}_j} \left( b_j^\dagger a_{\mathbf{q}_{\parallel},n}^{\phantom{\dagger}} - b_j a_{\mathbf{q}_{\parallel},n} \right) + \mathrm{H.c.}\right] \sin\left(\frac{\pi n}{L}z_j\right) \nonumber\\
        &+ \frac{2}{N}\sum_{j=1}^N \sum_{\mathbf{q}_{\parallel},\mathbf{q}_{\parallel}'} \sum_{n,n'=1}^{n_\mathrm{c}} \hbar D_{q_{\parallel},n;q_{\parallel}',n'}
        \left[\mathrm{e}^{\mathrm{i} (\mathbf{q}_{\parallel}-\mathbf{q}_{\parallel}') \cdot \bm{\rho}_j} a_{\mathbf{q}_{\parallel},n\phantom{'}}^{\phantom{\dagger}} a_{\mathbf{q}_{\parallel}',n'}^\dagger - \mathrm{e}^{\mathrm{i} (\mathbf{q}_{\parallel}+\mathbf{q}_{\parallel}') \cdot \bm{\rho}_j} a_{\mathbf{q}_{\parallel},n\phantom{'}} a_{\mathbf{q}_{\parallel}',n'} + \mathrm{H.c.} \right]  
        \nonumber\\
        &\times\sin\left(\frac{\pi n}{L}z_j\right) \sin\left(\frac{\pi n'}{L}z_j\right).\label{eq:bulk_inter_h} 
    \end{align}
    The coupling constants in the above equation read
    \begin{subequations}
        \begin{align}
            g_{q_{\parallel},n} &= \frac{\nu_\mathrm{0}}{2}\, \frac{\omega_{0,n}\sqrt{\omega_0}}{\omega_{q_{\parallel},n}^{3/2}} , \\ D_{q_{\parallel},n;q_{\parallel}',n'} &= \frac{g_{q_{\parallel},n\phantom{'}}g_{q_{\parallel}',n'}}{\omega_0}, 
        \end{align}  
    \end{subequations}
    where $\nu_0=(Q^2/M\varepsilon_0 d^3)^{1/2}$ is the ionic plasma frequency that characterizes the strength of the coupling of the ground-state vibrational modes to the cavity photons.
    Note that in Eq.~\eqref{eq:bulk_inter_h}, $n_\mathrm{c}=L/d$ is a cutoff that is imposed by our dipolar approximation.

    The perturbative part of the Hamiltonian $H$ writes 
    \begin{equation}
        \label{eq:H_1_pert}
        H_1 = \mathrm{i}\, \sqrt{\frac{2}{N}} \sum_{j=1}^N \sum_{\mathbf{q}_{\parallel}} \sum_{n=1}^{n_\mathrm{c}} \hbar \xi_{q_{\parallel},n} P_j^{(\mathrm{e})} \left(\mathrm{e}^{\mathrm{i} \mathbf{q}_{\parallel} \cdot \bm{\rho}_j} a_{\mathbf{q}_{\parallel},n} -  \mathrm{H.c.}\right) \sin\left(\frac{\pi n}{L}z_j\right),
    \end{equation}
    with 
    \begin{equation}
        \xi_{q_{\parallel},n} = \nu_\mathrm{e}\,\frac{\omega_{0,n}}{\omega_{q_{\parallel},n}} \sqrt{\frac{1}{2\hbar m_\mathrm{e} \omega_{q_{\parallel},n}}} .
    \end{equation}
    Here, $\nu_\mathrm{e}=(e^2/m_\mathrm{e} \varepsilon_0 d^3)^{1/2}$ is the electronic plasma frequency.

\subsection{Diagonalization of the unperturbed Hamiltonian}\label{subsec:bulk_cavity_diag}

\subsubsection{Collective vibrational operators}\label{subsubsec:collective_operators}
    In order to diagonalize the nonperturbative Hamiltonian \eqref{eq:bulk_inter_h},
    we introduce collective operators for the vibrational modes of the molecules, defined as [cf.\ Eq.~(2) of the main text]
    \begin{equation}
        \label{eq:S_operator}
        S_{\mathbf{k}_{\parallel},n} = \sqrt{\frac{2}{N}} \sum_{j=1}^N \mathrm{e}^{-\mathrm{i} \mathbf{k}_{\parallel} \cdot \bm{\rho}_j} \sin\left(\frac{\pi n}{L}z_j\right) b_j,
    \end{equation}
    with $\mathbf{k}_{\parallel} = k_x \hat{\mathbf{x}} + k_y \hat{\mathbf{y}}$ the in-plane wavevector. With the above definition, one has 
    $[S_{\mathbf{k}_{\parallel},n},S_{\mathbf{k}_{\parallel}',n'}] = [S^\dagger_{\mathbf{k}_{\parallel},n},S^\dagger_{\mathbf{k}_{\parallel}',n'}] = 0$, while 
    \begin{equation}
        \left[S_{\mathbf{k}_\parallel,n}^{\phantom{\dagger}},{S_{\mathbf{k}_\parallel',n'}^\dagger}\right] = 2 \left[ \frac{1}{N_\parallel} \sum_{\bm{\rho}_j} \mathrm{e}^{-\mathrm{i} (\mathbf{k}_\parallel - \mathbf{k}_\parallel') \cdot \bm{\rho}_{j}} \right] \left[ \frac{1}{N_z} \sum_{z_j} \sin\left(\frac{\pi n}{L} z_j\right) \sin\left(\frac{\pi n'}{L} z_{j}\right) \right].
    \end{equation}
    In the limit of a large number of sites ($N_\parallel,\,N_z \gg 1$), the first term in the square brackets can be simplified to the Kronecker delta $\delta_{\mathbf{k}_\parallel\mathbf{k}_\parallel'}$, while the second term can be rewritten as the integral
    \begin{equation}
        \frac{1}{L} \int_0^L \mathrm{d}z \sin\left(\frac{\pi n}{L} z\right) \sin\left(\frac{\pi n'}{L} z\right) = \frac{\delta_{nn'}}{2}.
    \end{equation}
    Therefore, in the thermodynamic limit the operator \eqref{eq:S_operator} satisfies the bosonic commutation relation $[S_{\mathbf{k}_\parallel,n}^{\phantom{\dagger}}{S_{\mathbf{k}_\parallel',n'}^\dagger}] = \delta_{\mathbf{k}_\parallel\mathbf{k}_\parallel'} \delta_{nn'}$.

    With the definition \eqref{eq:S_operator}, we have 
    \begin{equation}
        \sum_{\mathbf{k}_\parallel} \sum_{n=1}^{n_\mathrm{c}} S_{\mathbf{k}_\parallel,n}^\dagger S_{\mathbf{k}_\parallel,n}^{\phantom{\dagger}} = 2 \sum_{j,j'} \left[\frac{1}{N_\parallel} \sum_{\mathbf{k}_\parallel} \mathrm{e}^{-\mathrm{i} \mathbf{k}_\parallel \cdot (\bm{\rho}_j - \bm{\rho}_{j'})}\right]  \left[ \frac{1}{N_z} \sum_{n=1}^{n_\mathrm{c}} \sin\left(\frac{\pi n}{L} z_j\right) \sin\left(\frac{\pi n}{L} z_{j'}\right) \right] b_j^\dagger b_{j'}.
    \end{equation}
    The first term in the square brackets reads $\delta_{\bm{\rho}_j\bm{\rho}_{j'}}$. For the second term, we use that $z_j=(l-1)L/(N_z-1)\simeq (l-1)d$, where $l=1,\ldots,N_z$ labels the different molecular layers.
    We further define $\eta = {\pi nd}/{L} = \pi n/N_z$, such that $\Delta \eta = {\pi}/{N_z} \to 0$. Using the value of the cutoff $n_\mathrm{c} = L/d = N_z$, we replace the sum over $n$ by an integral over $\eta$ that reads
    \begin{equation}
        \frac{1}{\pi} \int_0^\pi \mathrm{d}\eta \sin(\eta [l-1]) \sin(\eta [l'-1]) = \frac{\delta_{ll'}}{2}.
    \end{equation}
    We thus obtain that 
    \begin{equation}
        \sum_{\mathbf{k}_\parallel} \sum_{n=1}^{n_\mathrm{c}} S_{\mathbf{k}_\parallel,n}^\dagger S_{\mathbf{k}_\parallel,n}^{\phantom{\dagger}}
        =\sum_{j=1}^N  b_j^\dagger b_j^{\phantom{\dagger}}.
    \end{equation}

    Finally, we identify the $S_{\mathbf{k}_\parallel,n}$ operators in Eq.~\eqref{eq:bulk_inter_h} and implement in-plane momentum conservation, so that we get $H^{(\mathrm{int})}_0 = \sum_{\mathbf{q}_\parallel} \sum_{n=1}^{n_\mathrm{c}} H^{(\mathrm{int})}_{\mathbf{q}_\parallel,n}$, where 
    \begin{align}
        \label{eq:bulk_int_h_mode}
        H^{(\mathrm{int})}_{\mathbf{q}_\parallel,n} =&\; \hbar \omega_{q_\parallel,n} a_{\mathbf{q}_\parallel,n}^\dagger a_{\mathbf{q}_\parallel,n}^{\phantom{\dagger}} + \hbar \omega_0 S_{\mathbf{q}_\parallel,n}^\dagger S_{\mathbf{q}_\parallel,n}^{\phantom{\dagger}} + \hbar g_{q_\parallel,n} \left( S_{\mathbf{q}_\parallel,n}^\dagger a_{\mathbf{q}_\parallel,n}^{\phantom{\dagger}} - S_{\mathbf{q}_\parallel,n} a_{-\mathbf{q}_\parallel,n} + \mathrm{H.c.} \right) \nonumber \\
        &+ \hbar D_{q_\parallel,n} \left( a_{\mathbf{q}_\parallel,n}^{\phantom{\dagger}} a_{\mathbf{q}_\parallel,n}^\dagger - a_{\mathbf{q}_\parallel,n} a_{-\mathbf{q}_\parallel,n} + \mathrm{H.c.} \right), 
    \end{align}
    which corresponds to Eq.~(3) of the main text.

\subsubsection{Hopfield-Bogoliubov diagonalization}
    For each mode $\{\mathbf{q}_\parallel,n\}$, we introduce the Hopfield-Bogoliubov operators defined as
    \begin{equation}
        p_{\mathbf{q}_\parallel,n}^{\pm} = w_{q_\parallel,n}^{\pm} a_{\mathbf{q}_\parallel,n} + x_{q_\parallel,n}^{\pm} S_{\mathbf{q}_\parallel,n} + y_{q_\parallel,n}^{\pm} a_{-\mathbf{q}_\parallel,n}^\dagger + z_{q_\parallel,n}^{\pm} S_{-\mathbf{q}_\parallel,n}^\dagger
    \end{equation}
    and impose the bosonic commutation relations 
    \begin{equation}
        \label{eq:p_commutator}
        \left[p_{\mathbf{q}_\parallel,n}^{\pm},{p_{\mathbf{q}_\parallel',n'}^{\pm\dagger}}\right] = \delta_{\mathbf{q}_\parallel\mathbf{q}_\parallel'} \delta_{nn'}.
    \end{equation}
    For these operators to diagonalize the Hamiltonian \eqref{eq:bulk_int_h_mode}, we need to solve the equation of motion $[p_{\mathbf{q}_\parallel,n}^{\pm},{H^{(\mathrm{int})}_{\mathbf{q}_\parallel,n}}] = \hbar \Omega^\pm_{q_\parallel,n} p_{\mathbf{q}_\parallel,n}^{\pm}$ \cite{hopfield_theory_1958}. This leads to the matrix
    \begin{equation}
        \label{eq:M_matrix}
        \mathcal{M}_{q_\parallel,n} =  \begin{pmatrix}
            \omega_{q_\parallel,n} + 2 D_{q_\parallel,n} &  g_{q_\parallel,n} & 2 D_{q_\parallel,n} & g_{q_\parallel,n}\\
            g_{q_\parallel,n} & \omega_0 & g_{q_\parallel,n} & 0\\
            -2 D_{q_\parallel,n} & -g_{q_\parallel,n} & -\omega_{q_\parallel,n} - 2 D_{q_\parallel,n} & -g_{q_\parallel,n}\\
            -g_{q_\parallel,n} & 0 & -g_{q_\parallel,n} & -\omega_0
        \end{pmatrix}
    \end{equation}
    solving $\mathcal{M}_{q_\parallel,n} ( w_{q_\parallel,n}, x_{q_\parallel,n}, y_{q_\parallel,n}, z_{q_\parallel,n} )^\mathrm{t} = \Omega^{\pm}_{q_\parallel,n} (w_{q_\parallel,n}, x_{q_\parallel,n}, y_{q_\parallel,n}, z_{q_\parallel,n})^\mathrm{t}$. Diagonalizing $\mathcal{M}_{q_\parallel,n}$, the two positive eigenvalues $\Omega^{\pm}_{q_\parallel,n}$ of Eq.~\eqref{eq:M_matrix} are found to be
        \begin{equation}
        \Omega^{\pm}_{q_\parallel,n} = {\omega_0}\,\sqrt{\frac{1 + \chi_{q_\parallel,n} + {( {\omega_{q_\parallel,n}}/{\omega_0} )}^2}{2} \pm \sqrt{\frac{{( {\omega_{q_\parallel,n}}/{\omega_0} )}^4 - 2 {( {\omega_{q_\parallel,n}}/{\omega_0} )}^2 (1 - \chi_{q_\parallel,n} ) + {(1 + \chi_{q_\parallel,n})}^2}{4}}},
    \end{equation}
    where 
    \begin{equation}
        \chi_{q_\parallel,n} = 4 \left(\frac{g_{q_\parallel,n}}{\omega_0}\right)^2 \frac{\omega_{q_\parallel,n}}{ \omega_0} = \left(\frac{\nu_0}{\omega_0}\, \frac{\omega_{0,n}}{\omega_{q_\parallel,n}}\right)^2.
    \end{equation}

    Using that $|w_{q_\parallel,\,n}|^2 + |x_{q_\parallel,\,n}|^2 - |y_{q_\parallel,\,n}|^2 - |z_{q_\parallel,\,n}|^2 = 1$ imposed by Eq.~\eqref{eq:p_commutator}, the eigenvectors of the matrix \eqref{eq:M_matrix} are given by the coefficients
    \begin{subequations}
        \begin{align}
            w_{q_\parallel,n}^{\pm} &= -\frac{(\omega_0 + \Omega^{\pm}_{q_\parallel,n})(\omega_{q_\parallel,n} + \Omega^{\pm}_{q_\parallel,n})}{2 \omega_{q_\parallel,n} g_{q_\parallel,n}} \,z_{q_\parallel,n}^{\pm},\\
            x_{q_\parallel,n}^{\pm} &= \frac{\omega_0 + \Omega^{\pm}_{q_\parallel,n}}{\omega_0 - \Omega^{\pm}_{q_\parallel,n}}\, z_{q_\parallel,n}^{\pm},\\
            y_{q_\parallel,n}^{\pm} &= -\frac{(\omega_0 + \Omega^{\pm}_{q_\parallel,n})(\omega_{q_\parallel,n} - \Omega^{\pm}_{q_\parallel,n})}{2 \omega_{q_\parallel,n} g_{q_\parallel,n}} \,z_{q_\parallel,n}^{\pm},\\
            z_{q_\parallel,n}^{\pm} &= \sqrt{\frac{\omega_{q_\parallel,n} g_{q_\parallel,n}^2 {(\omega_0 - \Omega^{\pm}_{q_\parallel,n})}^2}{\Omega^{\pm}_{q_\parallel,n} {[\omega_0^2 - \left(\Omega^{\pm}_{q_\parallel,n}\right)^2]}^2 + 4 \Omega^{\pm}_{q_\parallel,n} \omega_0 \omega_{q_\parallel,n} g_{q_\parallel,n}^2}}.
        \end{align}  
    \end{subequations}
    From these coefficients, we respectively define the light and matter weights of each polaritonic mode as
        $\mathrm{Ph}_{q_\parallel,n}^{\pm} = | w_{q_\parallel,n}^{\pm}|^2 - | y_{q_\parallel,n}^{\pm}|^2$ and
        $\mathrm{Mat}_{q_\parallel,n}^{\pm} = | x_{q_\parallel,n}^{\pm}|^2 -| z_{q_\parallel,n}^{\pm}|^2$,
    with their sum always equal to one.

\subsection{Raman scattering transition element}\label{subsec:raman_scattering}
    To compute the Raman scattering spectra we employ perturbation theory, treating the Hamiltonian $H_1$ [see Eq.~\eqref{eq:H_1_pert}] as a perturbation. We consider an initial photon with energy $\hbar \omega_\mathrm{L}$ impinging from a laser perpendicular to the cavity plane [see Fig.~1(a) of the main text]. We then look at the transition amplitude where the system emits a scattered photon at an energy $\hbar \omega_\mathrm{S}$ with an angle $\theta_\mathrm{S}$ to the $z$ axis and further creates a given $\pm$ polaritonic state into the system with in-plane wavevector $\mathbf{q}_\parallel$ and mode number $n$.
    In order to get the experimental Raman intensity of the system, we then must sum over all possible final polaritonic states. In principle, we should model the laser and scattered photonic states as polaritonic ones. However, as the laser is far detuned from the cavity modes ($\omega_\mathrm{L} \simeq  \omega_\mathrm{S} \gg \omega_0$), we can safely model such states as free photons. We denote the laser and scattered photon in-plane wavevectors as $\mathbf{q}_\parallel^\mathrm{L}$ and $\mathbf{q}_\parallel^\mathrm{S}$, respectively, and their mode numbers as $n_\mathrm{L}$ and $n_\mathrm{S}$.
     We restrict ourselves to Raman processes for which  $\mathbf{q}_\parallel^\mathrm{L} \neq \mathbf{q}_\parallel^\mathrm{S}$, and ignore Rayleigh scattering events.
    
    The initial and final states of the system can then be written as $\ket{\mathrm{I}} = a_{\mathbf{q}_\parallel^\mathrm{L},n_\mathrm{L}}^\dagger \ket{\mathrm{G}}$ and $\ket{\mathrm{F}} = p_{\mathbf{q}_\parallel,n}^{\pm\dagger} a_{\mathbf{q}_\parallel^\mathrm{S},n_\mathrm{S}}^\dagger \ket{\mathrm{G}}$, respectively, where $\ket{\mathrm{G}}$ is the ground state of the system. Such a choice corresponds to probing Stokes Raman processes, where the energy difference $\hbar \omega_\mathrm{L} - \hbar \omega_\mathrm{S}$ is transferred to the system. In the case where the vibrational coupling strength $\nu_0$ is not too large, the ground state of the system can be approximated by considering every molecule in its ground state manifold with no vibrational excitation and the cavity in its vacuum state, such that $\ket{\mathrm{G}} = \bigotimes_{j=1}^N \ket{{v}=0}_j \otimes \ket{0}$.
    This approximation is valid even at room temperature, as the vibrational energy $\hbar\omega_0 \simeq\unit[200]{meV}$ is much larger than the thermal energy $k_\mathrm{B} T \simeq \unit[25]{meV}$.

    Due to selection rules, the first nonvanishing order contribution to the Raman transition amplitude is of second order in $H_1$. The total transition amplitude from the initial to the final state is thus given by $\Gamma = \sum_{\sigma=\pm} \sum_{\mathbf{q}_\parallel} \sum_{n=1}^{n_\mathrm{c}} \Gamma^s_{\mathbf{q}_\parallel,n}$, where \cite{cohen-tannoudji_atom_1998}
    \begin{equation}
        \Gamma^\pm_{\mathbf{q}_\parallel,n} = \frac{2\pi}{\hbar} \delta\!\left(\hbar \omega_\mathrm{L} - \hbar \omega_\mathrm{S} - \hbar \Omega^{\pm}_{q_\parallel,n}\right) \left| \sum_\alpha 
        \frac{\langle \mathrm{F}|H_1|\alpha\rangle\langle\alpha|H_1|\mathrm{I}\rangle}{\hbar \omega_\mathrm{L} - \hbar \omega_\alpha + \mathrm{i}0^+} \right|^2.
    \end{equation}
    Here, the sum runs over all possible intermediate states $\ket{\alpha}$ of the system with energy $\hbar \omega_\alpha$. By replacing the different terms and computing the matrix elements, we find that the only intermediate states that contribute are those where one or several molecules are in the excited manifold, that is the states $\ket{w}_j$ of the Hamiltonian \eqref{eq:H_j_v1}. We restrict ourselves to the first excitation subspace in the electronic excited manifold, where only one molecule can be excited at a time, as it is the dominant term in the Raman process. Using Eq.~\eqref{eq:S_operator}, we obtain that the in-plane momentum is conserved, that is $\mathbf{q}_\parallel^{\phantom{\mathrm{S}}} = \mathbf{q}_\parallel^\mathrm{L} - \mathbf{q}_\parallel^\mathrm{S}$. 
    The total Raman scattering rate then reads 
    \begin{equation}
    \label{eq:Gamma_layerS}
        \Gamma = \gamma\sum_{\sigma=\pm} \sum_{n=1}^{n_\mathrm{c}} \delta\!\left(\omega_\mathrm{L} - \omega_\mathrm{S} - \Omega^s_{|\mathbf{q}_\parallel^\mathrm{L} - \mathbf{q}_\parallel^\mathrm{S}|,n}\right) \left( x^s_{|\mathbf{q}_\parallel^\mathrm{L} - \mathbf{q}_\parallel^\mathrm{S}|,n} \right)^2 f_{n},
    \end{equation}
    with
    \begin{equation}
        \gamma = \frac{2\pi}{N} \left[\sum_w \left(\xi_{q_\parallel^\mathrm{L},n_\mathrm{L}}\mu_0^w \right)\left(\xi_{q_\parallel^\mathrm{S},n_\mathrm{S}}\mu_1^w\right) \left(\frac{1}{\omega_\mathrm{L}-\omega_\mathrm{e}-\omega_w} - \frac{2}{\omega_\mathrm{S}+\omega_\mathrm{e}+\omega_w} \right)\right]^2,
    \end{equation}
    where $\mu_{v=0,1}^w=\langle v|_jP^{(\mathrm{e})}_j| w\rangle_j$ (assumed to be the same for every molecule $j$).
    In Eq.~\eqref{eq:Gamma_layerS}, 
    \begin{equation}
        f_n = \left[ \frac{2\sqrt{2}}{N_z} \sum_{z_j} \sin \left(\frac{\pi n_\mathrm{L}}{L} z_j\right) \sin \left(\frac{\pi n}{L} z_j\right)\sin \left(\frac{\pi n_\mathrm{S}}{L} z_{j}\right) \right]^2.
    \end{equation}
    Moving to the continuum limit [cf.\ Eq.\ (5) of the main text], using trigonometric identities allows us to show that
    \begin{equation}\label{eq:selection_rules}
        f_n = \frac{64}{\pi^2}\frac{(1-\delta_{n, n_\mathrm{L}\pm n_\mathrm{S}})[1 - (-1)^{n_\mathrm{L}+n+n_\mathrm{S}}] \left(n_\mathrm{L} n\, n_\mathrm{S}\right)^2}{{[(n_\mathrm{L} + n_\mathrm{S})^2 - n^2]}^2 {[(n_\mathrm{L} - n_\mathrm{S})^2 - n^2]}^2}.
    \end{equation}

\subsection{Reference case: incoherent molecules}\label{subsec:incoherent}
    Throughout the main text, the Raman spectra are normalized to the response of an \emph{incoherent} ensemble of molecules organized in a similar geometry as the coherent one, which also coincides with the \emph{nonresonant} situation in which the cavity modes are far detuned from the vibrational frequency $\omega_0$. We derive here this reference rate, using a similar approach as in Ref.~\cite{delpino_signatures_2015}.

    The collective operators $S_{\mathbf{k}_\parallel,n}$ of Eq.~\eqref{eq:S_operator} encode a fixed phase relation between the vibrational excitations of the different molecules. It is precisely this phase coherence, inherited from the cavity mode profile, that gives rise to the polaritonic coherent eigenstates and to the selection rule \eqref{eq:selection_rules}. In the incoherent (or nonresonant) case, no such coherent superposition is formed: Each molecule scatters independently, and the final state of a Raman process is a single localized vibrational excitation $b_j^\dagger \ket{\mathrm{G}}$ rather than a delocalized coherent polariton. Repeating the second-order perturbative calculation of Sec.~\ref{subsec:raman_scattering} with these uncorrelated final states, the transition amplitude no longer interferes constructively across molecules. This yields
    \begin{equation}\label{eq:Gamma_incoherent}
        \Gamma_\mathrm{inc} = \gamma\, \delta\!\left(\omega_\mathrm{L} - \omega_\mathrm{S} - \omega_0\right) f_\mathrm{inc},
    \end{equation}
    where the scaling factor $f_\mathrm{inc}$ is given by
    \begin{equation}\label{eq:f_incoherent}
        f_\mathrm{inc} = \frac{4}{N_z} \sum_{z_j} \sin^2 \left(\frac{\pi n_\mathrm{L}}{L} z_j\right) \sin^2 \left(\frac{\pi n_\mathrm{S}}{L} z_j\right).
    \end{equation}
    Equation~\eqref{eq:Gamma_incoherent} thus defines the reference height used to normalize all Raman spectra in the main text, with a varying scaling factor \eqref{eq:f_incoherent} that depends on the filling of the cavity.
    In order to further simplify the notations as well as allowing comparison between the incoherent and coherent cases, we can rewrite generally Eq.~\eqref{eq:Gamma_incoherent}, separating the peak position and height, by introducing the notation
    \begin{equation}
        \Gamma_\mathrm{inc} = \Gamma_0\, \delta\!\left([\omega_\mathrm{L} - \omega_\mathrm{S} - \omega_0]/{\omega_0}\right) 
    \end{equation}
        where
    \begin{equation}
        \label{eq:Gamma_incoherent_normalized}
        \quad\Gamma_0 = \frac{\gamma}{\omega_0} f_\mathrm{inc},
    \end{equation}
    such that $\Gamma_0$ is the incoherent scattering rate at the bare vibrational frequency $\omega_0$.
    Using the above notation, Eq.~\eqref{eq:Gamma_layerS} can be rewritten as a normalized scattering rate that reads
    \begin{equation}\label{eq:Gamma_normalized}
        \Gamma = \Gamma_0 \sum_{\sigma=\pm} \sum_{n=1}^{n_\mathrm{c}} \delta\!\left(\left[\omega_\mathrm{L} - \omega_\mathrm{S} - \Omega^s_{|\mathbf{q}_\parallel^\mathrm{L} - \mathbf{q}_\parallel^\mathrm{S}|,n}\right]/\omega_0\right) \left( x^s_{|\mathbf{q}_\parallel^\mathrm{L} - \mathbf{q}_\parallel^\mathrm{S}|,n} \right)^2 f_{n}.
    \end{equation}
    We thus recover the scattering rate $\Gamma$ from the main text [cf.\ Eq.\ (4)].

    In this geometry where the cavity is fully filled by molecules, Eq.~\eqref{eq:f_incoherent} can be simplified by moving the sum to an integral in the continuum limit,
    \begin{equation}
        f_\mathrm{inc} = \frac{4}{L} \int_0^L \mathrm{d}z \sin^2 \left(\frac{\pi n_\mathrm{L}}{L} z\right) \sin^2 \left(\frac{\pi n_\mathrm{S}}{L} z\right) = 1,
    \end{equation}
    such that Eq.~\eqref{eq:Gamma_incoherent} becomes $\Gamma_\mathrm{inc} = \gamma\, \delta\!\left(\omega_\mathrm{L} - \omega_\mathrm{S} - \omega_0\right)$.
    The incoherent ensemble thus produces a single Raman peak at the bare vibrational frequency $\omega_0$, with neither polaritonic splitting nor any dependence on the cavity-vibration spatial overlap: The selection-rule factor $f_n$ is absent, and the vibrational weight reduces to unity.

\section{Single molecular layer in a cavity}\label{sec:single_layer}
We consider in this section a single layer of $N$ molecules placed 
in the previous Fabry-Perot photonic cavity. The molecular layer is parallel to the cavity mirrors, and located 
at a distance $h$ with respect to the bottom mirror. For simplicity, the molecules are assumed to be arranged on a square lattice with lattice constant $d$.

\subsection{Hamiltonian and eigenvalues of the system}
    Following a similar derivation as in Sec.~\ref{sec:bulk_cavity}, we can write the total Hamiltonian of the system as a sum of a nonperturbative part $H_0 = H^\mathrm{{(int)}}_0 + H^{(\mathrm{v}1)}_0$ with $H^{(\mathrm{v}1)}_0$ given in Eq.~\eqref{eq:H_j_v1} and a perturbation $H_1$. We have 
    \begin{align}
        H^\mathrm{{(int)}}_0 &= \sum_{\mathbf{q}_\parallel} \sum_{n=1}^{n_\mathrm{c}} \hbar \omega_{q_\parallel,n} a_{\mathbf{q}_\parallel,n}^\dagger a_{\mathbf{q}_\parallel,n}^{\phantom{\dagger}}
        + \sum_{j=1}^N \hbar \omega_0 b_j^\dagger b_j^{\phantom{\dagger}}
        - \frac{1}{\sqrt{N}}\sum_{j=1}^N \sum_{\mathbf{q}_\parallel} \sum_{n=1}^{n_\mathrm{c}} \hbar \tilde g_{q_\parallel,n}  \left[\mathrm{e}^{\mathrm{i} \mathbf{q}_\parallel \cdot \bm{\rho}_j} \left( b_j^\dagger a_{\mathbf{q}_\parallel,n}^{\phantom{\dagger}} - b_j a_{\mathbf{q}_\parallel,n} \right) - \mathrm{H.c.}\right] \nonumber \\
        &\quad + \frac{1}{N}\sum_{j=1}^N \sum_{\mathbf{q}_\parallel,\mathbf{q}_\parallel'} \sum_{n,n'=1}^{n_\mathrm{c}} \hbar \tilde D_{q_\parallel,n;q_\parallel',n'} \left[\mathrm{e}^{\mathrm{i} (\mathbf{q}_\parallel-\mathbf{q}_\parallel') \cdot \bm{\rho}_j} a_{\mathbf{q}_\parallel,n}^{\phantom{\dagger}} a_{\mathbf{q}_\parallel',n'}^\dagger - \mathrm{e}^{\mathrm{i} (\mathbf{q}_\parallel+\mathbf{q}_\parallel') \cdot \bm{\rho}_j} a_{\mathbf{q}_\parallel,n} a_{\mathbf{q}_\parallel',n'} + \mathrm{H.c.} \right], \\
        H_1 &= \frac{\mathrm{i}}{\sqrt{N}} \sum_{j=1}^N \sum_{\mathbf{q}_\parallel} \sum_{n=1}^{n_\mathrm{c}} \hbar \tilde \xi_{q_\parallel,n} P_j^{(\mathrm{e})} \left(\mathrm{e}^{\mathrm{i} \mathbf{q}_\parallel \cdot \bm{\rho}_j} a_{\mathbf{q}_\parallel,n} -  \mathrm{H.c.}\right),
    \end{align}
    where the normalized coupling constants are
    \begin{subequations}
    \begin{align}
        \tilde g_{q_\parallel,n} &= \frac{\nu_0}{\sqrt{2}} \frac{\omega_{0,n}\sqrt{\omega_0}}{\omega_{q_\parallel,n}^{3/2}} \sin\left(\frac{\pi n}{L} h\right),
        \\
        \tilde D_{q_\parallel,n;q_\parallel',n'}&= \frac{\tilde g_{q_\parallel,n}\tilde g_{q_\parallel',n'}}{\omega_0},\\
        \tilde \xi_{q_\parallel,n} &= \nu_\mathrm{e}\, \frac{\omega_{0,n}}{\omega_{q_\parallel,n}} \sqrt{\frac{1}{\hbar m_\mathrm{e} \omega_{q_\parallel,\,n}}}\sin\left(\frac{\pi n}{L} h\right).
    \end{align}
    \end{subequations}
    We note that in contrast to the case where the cavity is filled with molecules, the sinus term is now a constant for a given mode number $n$ and height $h$, and can then enter in the definitions of the coupling constants. This allows us to simplify the expressions of the collective matter operators that no longer depend on the mode number $n$,  which is here not a good quantum number. We thus define these operators as
    \begin{equation}
    \label{eq:_S_operator_1layer}
        S_{\mathbf{k}_\parallel} = \frac{1}{\sqrt{N}} \sum_{j=1}^N \mathrm{e}^{-\mathrm{i} \mathbf{k}_\parallel \cdot \bm{\rho}_j} b_j,
    \end{equation}
    such that they verify the usual bosonic commutation relations. Using Eq.~\eqref{eq:_S_operator_1layer}, we have that $H^{(\mathrm{int})}_0 = \sum_{\mathbf{q}_\parallel} H^{(\mathrm{int})}_{\mathbf{q}_\parallel}$, where 
    \begin{align}
    \label{eq:H_single_layer_notdiag}
        H^{(\mathrm{int})}_{\mathbf{q}_\parallel} =&\; \hbar \omega_0 S_{\mathbf{q}_\parallel}^\dagger S_{\mathbf{q}_\parallel}^{\phantom{\dagger}} + \sum_{n=1}^{n_\mathrm{c}} \hbar \omega_{q_\parallel,n} a_{\mathbf{q}_\parallel,n}^\dagger a_{\mathbf{q}_\parallel,n}^{\phantom{\dagger}}
        + \sum_{n=1}^{n_\mathrm{c}} \hbar \tilde g_{q_\parallel,n}  \left(S_{\mathbf{q}_\parallel}^\dagger a_{\mathbf{q}_\parallel,n}^{\phantom{\dagger}} - S_{\mathbf{q}_\parallel} a_{-\mathbf{q}_\parallel,n} + \mathrm{H.c.}\right)\nonumber \\
        +& \sum_{n,n'=1}^{n_\mathrm{c}} \hbar \tilde D_{q_\parallel,n;q_\parallel,n'} \left( a_{\mathbf{q}_\parallel,n}^{\phantom{\dagger}} a_{\mathbf{q}_\parallel,n'}^\dagger - a_{\mathbf{q}_\parallel,n} a_{-\mathbf{q}_\parallel,n'} + \mathrm{H.c.} \right).
    \end{align}
    
    As compared to the filled vacity case (Sec.~\ref{sec:bulk_cavity}), the Hamiltonian \eqref{eq:H_single_layer_notdiag} is no longer diagonal in the $n$ indices, so that we do no longer obtain two polaritonic branches per mode $\{\bm{q}_\parallel,n\}$, but instead obtain $n_\mathrm{c}+1$ branches. We label them using the index $\eta = 1, \ldots, n_\mathrm{c}+1$. We next introduce the bosonic Hopfield-Bogoliubov operators \cite{hopfield_theory_1958} 
    \begin{equation}
        p_{\mathbf{q}_\parallel}^\eta = \sum_{n=1}^{n_\mathrm{c}} w_{q_\parallel,n}^\eta a_{\mathbf{q}_\parallel,n} + x_{q_\parallel}^\eta S_{\mathbf{q}_\parallel} + \sum_{n=1}^{n_\mathrm{c}} y_{q_\parallel,n}^\eta a_{-\mathbf{q}_\parallel,n}^\dagger + z_{q_\parallel}^\eta S_{-\mathbf{q}_\parallel}^\dagger.
    \end{equation}
    The equation of motion $[p_{\mathbf{q}_\parallel}^\eta, H^{(\mathrm{int})}_{\mathbf{q}_\parallel}] = \hbar \tilde\Omega_{q_\parallel}^\eta p_{\mathbf{q}_\parallel}^\eta$ then leads to the eigenvalue problem $\mathcal{M}_{q_\parallel} \psi_{\mathbf{q}_\parallel} = \tilde\Omega_{q_\parallel}^\eta \psi_{\mathbf{q}_\parallel}$, where $\psi_{\mathbf{q}_\parallel} = (a_{\mathbf{q}_\parallel,1}, \ldots, a_{\mathbf{q}_\parallel,n_\mathrm{c}}, S_{\mathbf{q}_\parallel}, a_{-\mathbf{q}_\parallel,1}^\dagger, \ldots, a_{-\mathbf{q}_\parallel,n_\mathrm{c}}^\dagger, S_{-\mathbf{q}_\parallel}^\dagger)^\mathrm{t}$.
    The Hopfield-Bogoliubov matrix $\mathcal{M}_{q_\parallel}$ has dimension $[2(n_\mathrm{c}+1)] \times [2(n_\mathrm{c}+1)]$ written by blocks as
    \begin{equation}
        \label{eq:M_matrix_single_layer}
        \mathcal{M}_{q_\parallel} = \begin{pmatrix}
            \mathcal{D}_{q_\parallel} + \mathcal{G}_{q_\parallel} & \mathcal{G}_{q_\parallel} \\
            -\mathcal{G}_{q_\parallel} & -\mathcal{D}_{q_\parallel} - \mathcal{G}_{q_\parallel}
        \end{pmatrix},
    \end{equation}
    with 
    \begin{subequations}
        \begin{align}
            \mathcal{D}_{q_\parallel} &= \mathrm{Diag}\left(\omega_{q_\parallel,1}, \ldots, \omega_{q_\parallel,n_\mathrm{c}}, \omega_0\right),\\ 
            \mathcal{G}_{q_\parallel} &= \begin{pmatrix}
                \tilde D_{q_\parallel,1;q_\parallel, 1} & \ldots & \tilde D_{q_\parallel,n_\mathrm{c};q_\parallel, 1} & \tilde g_{q_\parallel,1} \\
                \vdots & \ddots & \vdots & \vdots \\
                \tilde D_{q_\parallel,1;q_\parallel, n_\mathrm{c}} & \ldots & \tilde D_{q_\parallel,n_\mathrm{c};q_\parallel,n_\mathrm{c}} & \tilde g_{q_\parallel,n_\mathrm{c}} \\
                \tilde g_{q_\parallel,1} & \ldots & \tilde g_{q_\parallel,n_\mathrm{c}} & 0
            \end{pmatrix}.
        \end{align}
    \end{subequations}
    
    We diagonalize the matrix \eqref{eq:M_matrix_single_layer} numerically to obtain the eigenvalues and the corresponding orthonormal eigenvector basis.
    As the matter weight of the eigenstates rapidly decreases when increasing $\eta$, we can safely restrict ourselves to the first few polaritonic branches in the numerical computation of the Raman scattering (see below).

\subsection{Raman scattering rate}
    To compute the Raman scattering rate for the single molecular layer in a Fabry-Perot cavity, we proceed analogously to Sec.~\ref{subsec:raman_scattering}, but adapt the notation to the present geometry. The initial state is given by $\ket{\mathrm{I}} = a_{\mathbf{q}_\parallel^\mathrm{L},n_\mathrm{L}}^\dagger \ket{\mathrm{G}}$ and the final state by $\ket{\mathrm{F}} = p_{\mathbf{q}_\parallel}^{\eta\dagger} a_{\mathbf{q}_\parallel^\mathrm{S},n_\mathrm{S}}^\dagger \ket{\mathrm{G}}$, where $\ket{\mathrm{G}}$ denotes the ground state. Conservation of in-plane momentum requires that $\mathbf{q}_\parallel = \mathbf{q}_\parallel^\mathrm{L} - \mathbf{q}_\parallel^\mathrm{S}$.
    The normalized transition rate for the Raman process is then given by
    \begin{equation}
        \tilde\Gamma = \tilde\Gamma_0\, \sum_{\eta=1}^{n_\mathrm{c}+1} \delta\!\left(\left[\omega_\mathrm{L} - \omega_\mathrm{S} - \tilde\Omega^\eta_{|\mathbf{q}_\parallel^\mathrm{L} - \mathbf{q}_\parallel^\mathrm{S}|} \right]/\omega_{0}\right) \left( x^\eta_{|\mathbf{q}_\parallel^\mathrm{L} - \mathbf{q}_\parallel^\mathrm{S}|} \right)^2,
    \end{equation}
    where $\tilde\Gamma_0$ is a prefactor analogous to the filled cavity case [see Eq.~\eqref{eq:Gamma_incoherent_normalized}].
    However, unlike the filled cavity scenario, the scaling factor $\tilde{f}_\mathrm{inc}$ [see Eq.~\eqref{eq:f_incoherent}] contained in $\tilde\Gamma_0$ is not the same as in the bulk incoherent case, since the molecules are now organized in a single layer rather than being homogeneously distributed across the cavity. This leads to a different spatial overlap between the cavity modes and the molecular layer, which is reflected in
    \begin{equation}
        \tilde{f}_\mathrm{inc} = 4 \sin^2\left(\frac{\pi n_\mathrm{L}}{L} h\right) \sin^2\left(\frac{\pi n_\mathrm{S}}{L} h\right).
    \end{equation}

\section{Case of intermediate cavity filling}\label{sec:intermediate_filling}
    In this section, we discuss the evolution of the selection rules for spontaneous Raman scattering when varying the thickness of the molecular medium inside the Fabry-Perot cavity. We consider the case of a cavity partially filled by a set of $N_z$ molecular layers along the cavity axis $z$ with interlayer distance $d$, such that the total thickness of the filled region is $\ell = N_z d \leqslant L$.
    The molecular medium is assumed to be homogeneous over the filled region.
    This configuration allows us to interpolate between the two limiting cases studied in Secs.~\ref{sec:bulk_cavity} and \ref{sec:single_layer}, corresponding respectively to $\ell = L$ and $\ell \ll L$.

\subsection{Generalization of the Hamiltonian}
    Following the same procedure as in Sec.~\ref{sec:bulk_cavity}, we obtain a similar Hamiltonian as in Sec.~\ref{subsec:bulk_cavity_full_h}.
    To simplify the following analysis, we restrict ourselves to the regime of small to moderate vibrational coupling strengths $\nu_0 \lesssim \omega_0$, such that the diamagnetic term can be neglected and the rotating wave approximation can be applied.
    Under these approximations, the interaction Hamiltonian \eqref{eq:bulk_inter_h} reads
    \begin{align}\label{eq:H_int_intermediate}
        H_0^{(\mathrm{int})} = \sum_{\mathbf{q}_{\parallel}} \sum_{n=1}^{n_\mathrm{c}} \hbar \omega_{q_{\parallel},n} a_{\mathbf{q}_{\parallel},n}^\dagger a_{\mathbf{q}_{\parallel},n}^{\phantom{\dagger}}
        + \sum_{j=1}^N \hbar \omega_0 b_j^\dagger b_j^{\phantom{\dagger}} + \sqrt{\frac{2}{N}}\sum_{j=1}^N \sum_{\mathbf{q}_{\parallel}} \sum_{n=1}^{n_\mathrm{c}} \hbar g_{q_{\parallel},n} \left(\mathrm{e}^{\mathrm{i} \mathbf{q}_{\parallel} \cdot \bm{\rho}_j} b_j^\dagger a_{\mathbf{q}_{\parallel},n}^{\phantom{\dagger}} + \mathrm{H.c.}\right) \sin\left(\frac{\pi n}{L}z_j\right).
    \end{align}
    Here, the sum over $j$ only runs over all molecules located within the filled region of thickness $\ell$ and can no longer be replaced by an integral over the full cavity length $L$.

    We introduce the collective matter operators along the in-plane wavevector $\mathbf{k}_\parallel$ as
    \begin{equation}
        S_{\mathbf{k}_\parallel,z_j} = \frac{1}{\sqrt{N_\parallel}} \sum_{\bm{\rho}_j} \mathrm{e}^{-\mathrm{i} \mathbf{k}_\parallel \cdot \bm{\rho}_j}\, b_j,
    \end{equation}
    which follow the usual bosonic commutation relations 
    $[S_{\mathbf{k}_\parallel,z_j}, S_{\mathbf{k}_\parallel',z_{j'}}^\dagger] = \delta_{\mathbf{k}_\parallel, \mathbf{k}_\parallel'} \delta_{z_j, z_{j'}}$
    and $[S_{\mathbf{k}_\parallel,z_j}, S_{\mathbf{k}_\parallel',z_{j'}}] = 0$.
    Using these operators, we can rewrite Eq.~\eqref{eq:H_int_intermediate} as a sum over independent in-plane wavevectors $\mathbf{q}_\parallel= \mathbf{k}_\parallel$ as $H_0 = \sum_{\mathbf{q}_\parallel} H^{(\mathrm{int})}_{\mathbf{q}_\parallel}$, where
    \begin{align}\label{eq:H_int_intermediate_q}
        H^{(\mathrm{int})}_{\mathbf{q}_\parallel} = \sum_{n=1}^{n_\mathrm{c}} \hbar \omega_{q_{\parallel},n} a_{\mathbf{q}_{\parallel},n}^\dagger a_{\mathbf{q}_{\parallel},n}^{\phantom{\dagger}}
        + \sum_{z_j} \hbar \omega_0 S_{\mathbf{q}_\parallel,z_j}^\dagger S_{\mathbf{q}_\parallel,z_j}^{\phantom{\dagger}}
        + \sqrt{\frac{2}{N_z}} \sum_{z_j} \sum_{n=1}^{n_\mathrm{c}} \hbar g_{q_{\parallel},n} \left(S_{\mathbf{q}_\parallel,z_j}^\dagger a_{\mathbf{q}_{\parallel},n}^{\phantom{\dagger}} + \mathrm{H.c.}\right) \sin\left(\frac{\pi n}{L} z_j\right).
    \end{align}

\subsection{Eigenvalues of the system}
    Formally, the eigenstates of Eq.~\eqref{eq:H_int_intermediate_q} can be written as a linear combination of the photonic and matter operators as
    \begin{equation}
        p_{\mathbf{q}_\parallel}^{\sigma} = \sum_{n=1}^{n_\mathrm{c}} w_{q_\parallel,n}^{\sigma} a_{\mathbf{q}_\parallel,n} + \sum_{z_j} x_{q_\parallel,z_j}^{\sigma} S_{\mathbf{q}_\parallel,z_j},
    \end{equation}
    where $\sigma = 1, \ldots, n_\mathrm{c} + N_z$ labels the polaritonic branches. We denote their eigenfrequencies as $\Omega_{q_\parallel}^\sigma$.
    The equation of motion $[p_{\mathbf{q}_\parallel}^\sigma, H^{(\mathrm{int})}_{\mathbf{q}_\parallel}] = \hbar \Omega_{q_\parallel}^\sigma p_{\mathbf{q}_\parallel}^\sigma$ leads to the eigenvalue problem $\mathcal{N}_{q_\parallel} \Psi_{\mathbf{q}_\parallel} = \Omega_{q_\parallel}^\sigma \Psi_{\mathbf{q}_\parallel}$, where 
    \begin{equation}
    \Psi_{\mathbf{q}_\parallel} = (a_{\mathbf{q}_\parallel,1}, \ldots, a_{\mathbf{q}_\parallel,n_\mathrm{c}}, S_{\mathbf{q}_\parallel,z_1}, \ldots, S_{\mathbf{q}_\parallel,z_{N_z}})^\mathrm{t}.
    \end{equation}
    The Hopfield-Bogoliubov matrix $\mathcal{N}_{q_\parallel}$ has dimension $(n_\mathrm{c} + N_z) \times (n_\mathrm{c} + N_z)$ and is written by blocks as
    \begin{equation}
    \label{eq:N_matrix}
        \mathcal{N}_{q_\parallel} = \begin{pmatrix}
            \mathrm{Diag}(\omega_{q_\parallel,1}, \ldots, \omega_{q_\parallel,n_\mathrm{c}}) & G_{q_\parallel} \\
            G_{q_\parallel}^\mathrm{t} & \omega_0 \mathbb{I}_{N_z}
        \end{pmatrix},
    \end{equation}
    where $\mathbb{I}_{N_z}$ is the identity matrix of dimension $N_z$ and the coupling matrix $G_{q_\parallel}$ has elements
    \begin{equation}
        (G_{q_\parallel})_{n,j} = \sqrt{\frac{2}{N_z}} g_{q_\parallel,n} \sin\left(\frac{\pi n}{L} z_j\right).
    \end{equation}

    \begin{figure}[tb]
        \centering
        \includegraphics[width=0.9\textwidth]{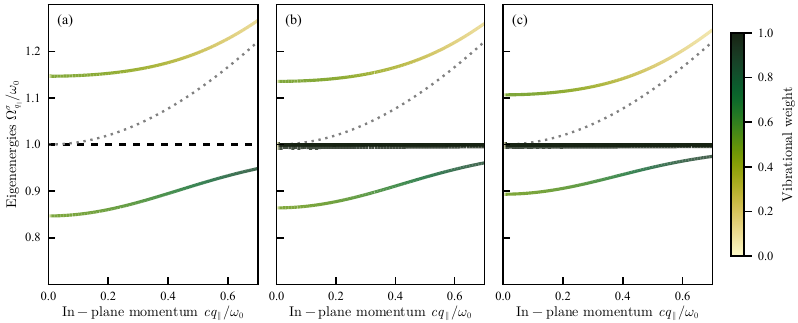}
        \caption{Polaritonic eigenfrequencies $\Omega_{q_\parallel}^\sigma$ as a function of the in-plane wavevector $q_\parallel$ for different molecular layer thicknesses $\ell$, all centered in the middle of the cavity: (a) $\ell \ll L$, (b) $\ell = 0.5 L$, and (c) $\ell = L$.
        The color scale indicates the vibrational matter weight $\sum_{z_j} |x_{q_\parallel,z_j}^\sigma|^2$ of each eigenstate.
        Parameters: $\nu_0 = 0.2\,\omega_0$, $L = \lambda_0/2$, $\lambda_0 = 800 d$, and $n_\mathrm{c} = 10$, where $\lambda_0 = 2\pi c/\omega_0$.}
        \label{fig:intermediate_filling}
    \end{figure}

    We diagonalize the matrix \eqref{eq:N_matrix} numerically to obtain the eigenvalues and the corresponding orthonormal eigenvector basis, as shown in Fig.~\ref{fig:intermediate_filling} for different filling fractions $\ell/L$ centered in the middle of the cavity. In the limit of a small filling fraction $\ell \ll L$ [Fig.~\ref{fig:intermediate_filling}(a)], we recover the results of Sec.~\ref{sec:single_layer}, where $N_z$ polaritonic branches appear due to the presence of $N_z$ molecular layers in the cavity, previously labeled by the index $\sigma = \eta$.
    When increasing the filling fraction $\ell/L$ [Fig.~\ref{fig:intermediate_filling}(b)], more polaritonic branches appear near the bare vibrational frequency $\omega_0$, as more molecular layers are present in the cavity.
    In the limit of a fully filled cavity $\ell = L$ [Fig.~\ref{fig:intermediate_filling}(c)], we recover the results of Sec.~\ref{sec:bulk_cavity}, where only two polaritonic branches appear for each cavity mode number $n$, labeled by the index $\sigma = (\pm, n)$.

    \begin{figure}[tb]
        \centering
        \includegraphics[width=0.9\textwidth]{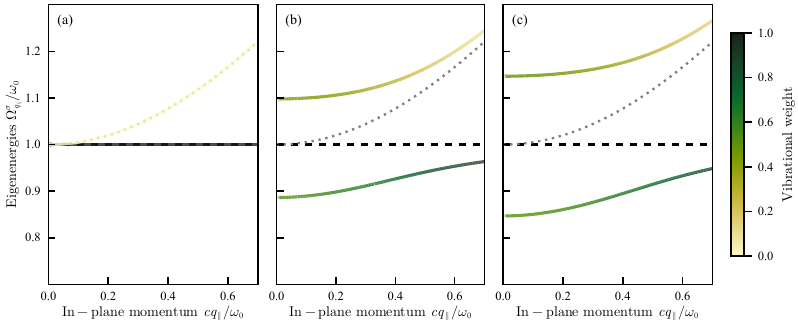}
        \caption{Similar figure as Fig.~\ref{fig:intermediate_filling} but for one single molecular layer located at different positions inside the cavity: (a) $h = 0$, (b) $h = 0.25 L$, and (c) $h = 0.5 L$.}
        \label{fig:intermediate_filling_single}
    \end{figure}

    In Fig.~\ref{fig:intermediate_filling_single}, we perform the same analysis as in Fig.~\ref{fig:intermediate_filling} but for one single molecular layer located at different positions inside the cavity. We find that the position of the molecular layer has a strong impact on the polaritonic eigenfrequencies and matter weights. For example, when the molecular layer is located at the center of the cavity, the coupling to the cavity modes is maximum as the layer is located at the antinodes of the cavity field, leading to a large Rabi splitting between the two polaritonic branches. In contrast, when the molecular layer is located close to one of the cavity mirrors, the coupling to the cavity modes is strongly reduced as the layer is located at the nodes of the cavity field, leading to a small Rabi splitting between the two polaritonic branches.

\subsection{Raman selection rule}
    Following the same procedure as in Sec.~\ref{subsec:raman_scattering}, we compute the Raman scattering rate for the intermediate filling case.
    The initial state is given by $\ket{\mathrm{I}} = a_{\mathbf{q}_\parallel^\mathrm{L},n_\mathrm{L}}^\dagger \ket{\mathrm{G}}$ and the final state by $\ket{\mathrm{F}} = p_{\mathbf{q}_\parallel}^{\sigma\dagger} a_{\mathbf{q}_\parallel^\mathrm{S},n_\mathrm{S}}^\dagger \ket{\mathrm{G}}$, where $\ket{\mathrm{G}}$ denotes the ground state. Conservation of in-plane momentum requires that $\mathbf{q}_\parallel = \mathbf{q}_\parallel^\mathrm{L} - \mathbf{q}_\parallel^\mathrm{S}$.
    The transition rate for the Raman process is then given by
    \begin{equation}\label{eq:Gamma_intermediate}
        \bar \Gamma = \bar\Gamma_0\, \sum_{\sigma=1}^{n_\mathrm{c}+N_z} \delta\!\left(\left[\omega_\mathrm{L} - \omega_\mathrm{S} - \Omega^\sigma_{|\mathbf{q}_\parallel^\mathrm{L} - \mathbf{q}_\parallel^\mathrm{S}|}\right]/\omega_{0}\right) |f^\sigma|^2,
    \end{equation}
    where $\bar\Gamma_0$ is the same prefactor as in Eq.~\eqref{eq:Gamma_incoherent_normalized} but with a different scaling factor $\bar{f}_\mathrm{inc}$, as the molecules are now homogeneously distributed over a finite region of thickness $\ell$ rather than filling the whole cavity [see Eq.~\eqref{eq:f_incoherent}]. This scaling factor is numerically computed for each filling fraction $\ell/L$ and position of the filled region inside the cavity. Lastly, the term $f^\sigma$ in Eq.~\eqref{eq:Gamma_intermediate} encodes the selection rule associated with the cavity mode indices for the intermediate filling configurations. It is given by
    \begin{equation}\label{eq:f_intermediate_original}
        f^\sigma = \sum_{z_j} x_{|\mathbf{q}_\parallel^\mathrm{L} - \mathbf{q}_\parallel^\mathrm{S}|,z_j}^\sigma \sin\left(\frac{\pi n_\mathrm{L}}{L} z_j\right) \sin\left(\frac{\pi n_\mathrm{S}}{L} z_j\right).
    \end{equation}
    As discussed in the main text, for $\nu/2\omega_0$ not too large, $n_\mathrm{L} - n_\mathrm{S} \simeq 1$.
    The selection rule \eqref{eq:f_intermediate_original} can then be rewritten as a function of $n_\mathrm{L}$,
    \begin{equation}\label{eq:f_intermediate}
        f^\sigma \simeq \sum_{z_j} x_{|\mathbf{q}_\parallel^\mathrm{L} - \mathbf{q}_\parallel^\mathrm{S}|,z_j}^\sigma \sin\left(\frac{\pi n_\mathrm{L}}{L} z_j\right) \sin\left(\frac{\pi (n_\mathrm{L} - 1)}{L} z_j\right).
    \end{equation}

    \begin{figure}[tb]
        \centering
        \includegraphics[width=\textwidth]{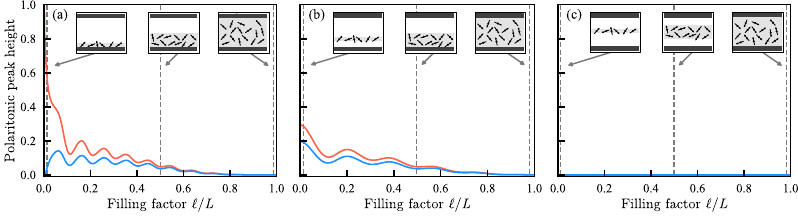}
        \includegraphics[width=\textwidth]{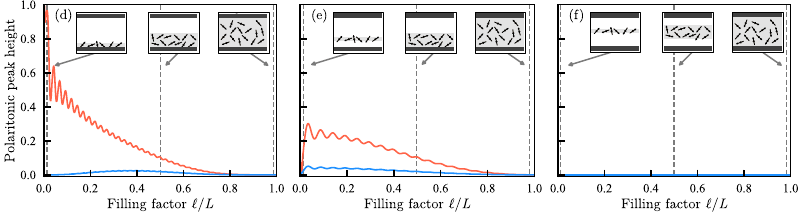}
        \caption{Height of the peaks relative to the bare uncoupled and incoherent case for the two resonant polaritonic branches $\sigma = \pm$ as a function of the cavity filling fraction $\ell/L$ for different positions of the filled region inside the cavity, centered at (a,d) $h=0$, (b,e) $L/4$, and (c,f) $L/2$.
        The inset of each panel schematically depicts the cavity for a given filling fraction $\ell/L$ and position of the filled region within the cavity.
        We consider similar parameters as in Fig.~\ref{fig:intermediate_filling} with $\omega_\mathrm{L} = 10.9\,\omega_0$ (a,b,c) and $\omega_\mathrm{L} = 40.0\,\omega_0$ (d,e,f), and $cq_\parallel / \omega_0 \simeq 0.01$.
        The red curve corresponds to the upper resonant polariton ($\sigma=+$) while the blue one to the lower resonant polariton ($\sigma=-$).}
        \label{fig:selection_rule_intermediate_filling}
    \end{figure}
    
    We then apply the following numerical procedure to analyze the evolution of the selection rule when varying the filling fraction $\ell/L$:
    \begin{itemize}
        \item For a given filling fraction $\ell/L$ centered at multiple positions in the cavity, we diagonalize the Hopfield-Bogoliubov matrix \eqref{eq:N_matrix} to obtain the eigenvalues $\Omega_{q_\parallel}^\sigma$ and the corresponding eigenvectors for a given in-plane wavevector $q_\parallel$.
        \item We identify the two polaritonic branches $\sigma = \pm$ that display the highest hybridization around the bare vibrational frequency $\omega_0$.
        \item We compute the selection rule $f^{\pm}$ using Eq.~\eqref{eq:f_intermediate}, the corresponding matter component $x_{q_\parallel,z_j}^{\pm}$ of the eigenvectors, as well as the scaling factor $\bar{f}_\mathrm{inc}$ to obtain the normalized height of the peaks associated with the two resonant polaritonic branches $\sigma = \pm$ in the Raman spectrum.
    \end{itemize}
    Our results for the height of the polaritonic peaks as a function of $\ell/L$ are displayed in Fig.~\ref{fig:selection_rule_intermediate_filling}.
    The data display rapid oscillations arising from the periodic spatial dependence of the sine functions in Eq.~\eqref{eq:f_intermediate}. 
    These oscillations are not significant as the overall trend of the selection rule remains robust when varying the laser frequency $\omega_\mathrm{L}$, provided that $n_\mathrm{L} - n_\mathrm{S} \simeq 1$.
    However, it is worth noting that the frequency of the laser $\omega_\mathrm{L}$ (and thus the mode number $n_\mathrm{L}$) can be used as an additional tuning parameter that also changes the height of the peaks in the Raman spectrum, as it changes the selection rule $f^\sigma$ through the sine functions in Eq.~\eqref{eq:f_intermediate}.

    Analyzing Fig.~\ref{fig:selection_rule_intermediate_filling} then demonstrates that the limiting cases discussed in Secs.~\ref{sec:bulk_cavity} and \ref{sec:single_layer} are consistently recovered, whathever the position of the filled region inside the cavity.
    When the cavity is fully filled, $f^{\sigma_\pm} = 0$, indicating that resonant polaritons cannot be observed. In contrast, for a single molecular layer, $f^{\sigma_\pm} \neq 0$, allowing both polaritonic branches to be detected in Raman scattering.
    Furthermore, when the molecular layer is positioned exactly at the center of the cavity, the selection rule always yields zero due to the symmetry of the cavity modes along the $z$ direction. This behavior can be understood analytically by examining the periodicity of the sine functions in Eq.~\eqref{eq:f_intermediate}.

    For intermediate filling fractions, the selection rule for the two resonant polaritonic branches $\sigma = \pm$ shows a nontrivial dependence on the filling fraction $\ell/L$ and the position of the filled region inside the cavity.
    Looking at the average trend of the selection rule when varying the filling fraction $\ell/L$, we find that the height of the peaks associated with the two resonant polaritonic branches $\sigma = \pm$ in the Raman spectrum is maximum in the limit of a small filling fraction $\ell \ll L$, and decreases when increasing the filling fraction $\ell/L$ until it vanishes in the limit of a fully filled cavity $\ell = L$.

    In order to optimally observe in an experiment both polaritonic peaks in the Raman spectrum, one should thus optimize the filling fraction and the position of the filled region inside the cavity to maximize the height of the two resonant polaritonic peaks. By looking at Fig.~\ref{fig:selection_rule_intermediate_filling}, we find that the optimal filling fraction is when $\ell \ll L$ for the parameters considered in this work, and that the filled region should be centered near $h \simeq 0$ or $h \simeq L$ in our configuration.

\section{Raman scattering of vibrational polaritons in free space}\label{sec:fs}
    To benchmark our model presented in Secs.~\ref{sec:bulk_cavity} and \ref{sec:single_layer}, we here compute the Raman scattering rate for a macroscopic ensemble of molecules in free space, i.e., without a Fabry-Perot cavity, assuming a spatial coherence extending over the quantization volume $V\rightarrow\infty$.
    We then compare our theoretical results with those obtained in the seminal experimental work by Henry and Hopfield on ionic crystals~\cite{henry_raman_1965}.

\subsection{Hamiltonian and its  diagonalization}
    We here consider a collection of $N\gg1$ Raman active molecules that are coupled to the vacuum electromagnetic modes.  
        We follow a similar procedure as detailed in the previous two sections. The free-space light modes (in a quantization volume $V\rightarrow\infty$) are now described by plane waves with three-dimensional wavevector $\mathbf{q}$
        and corresponding vector potential 
        \begin{equation}
            \mathbf{A}(\mathbf{r}) = \sum_{\mathbf{q}} \sum_{\nu=1,\,2} \sqrt{\frac{\hbar}{2\varepsilon_0 \omega_q V}} \left[\mathbf{u}^{\nu}_{\mathbf{q}}(\mathbf{r})\, a_{\mathbf{q}}^{\nu} + \mathbf{u}^{\nu *}_{\mathbf{q}}(\mathbf{r})\, a_{\mathbf{q}}^{\nu\dagger}\right], 
        \end{equation}
    where $\omega_q=cq$, $a_{\mathbf{q}}^{\nu}$ an operator which annihilates a photon with wavevector $\mathbf{q}$ and polarization $\nu$, and 
    \begin{subequations}
        \begin{align}
            \mathbf{u}_{\mathbf{q}}^1(\mathbf{r}) &= \mathrm{e}^{\mathrm{i} \mathbf{q} \cdot \mathbf{r}} \left[\cos \theta_{\mathbf{q}} \left(\cos \varphi_{\mathbf{q}} \hat{\mathbf{x}} + \sin \varphi_{\mathbf{q}} \hat{\mathbf{y}} \right) - \sin \theta_{\mathbf{q}} \hat{\mathbf{z}} \right],  \\
            \mathbf{u}_{\mathbf{q}}^2(\mathbf{r}) &= \mathrm{e}^{\mathrm{i} \mathbf{q} \cdot \mathbf{r}}  \left( \cos \varphi_{\mathbf{q}} \hat{\mathbf{y}} - \sin \varphi_{\mathbf{q}} \hat{\mathbf{x}} \right).
        \end{align}
    \end{subequations}
    We choose the vibrational and electronic dipoles of every molecule aligned in the $\hat{\mathbf{z}}$ direction. Similarly to the approximations performed in Sec.~\ref{subsec:bulk_cavity_h}, we restrict ourselves to the polarization $\nu = 1$ (dropping in what follows the corresponding index) and take $\varphi_{\mathbf{q}} = 0$. Using the same modelization of the molecules as in Sec.~\ref{subsec:molecules_desc}, we can write the total Hamiltonian of the system as a sum of a nonperturbative part $H_0 = H^\mathrm{{(int)}}_0 + H^{(\mathrm{v}1)}_0$ and a perturbation $H_1$. Here, $H^{(\mathrm{v}1)}_0$ is given in Eq.~\eqref{eq:H_j_v1} while
    \begin{align}
        H^\mathrm{{(int)}}_0 &= \sum_{\mathbf{q}} \hbar \omega_q a_{\mathbf{q}}^\dagger a_{\mathbf{q}}^{\phantom{\dagger}}
        + \sum_{j=1}^N \hbar \omega_0 b_j^\dagger b_j^{\phantom{\dagger}} - \frac{\mathrm{i}}{\sqrt{N}}\sum_{j=1}^N \sum_{\mathbf{q}} \hbar g_{\mathbf{q}}  \left[\mathrm{e}^{\mathrm{i} \mathbf{q} \cdot \mathbf{r}_j} \left( b_j^\dagger a_{\mathbf{q}}^{\phantom{\dagger}} - b_j a_{\mathbf{q}} \right) - \mathrm{H.c.}\right] \nonumber \\
        &\quad+ \frac{1}{N}\sum_{j=1}^N \sum_{\mathbf{q}\mathbf{q}'} \hbar D_{\mathbf{q}\mathbf{q}'}
        \left[\mathrm{e}^{\mathrm{i} (\mathbf{q}-\mathbf{q}') \cdot \mathbf{r}_j} a_{\mathbf{q}}^{\phantom{\dagger}} a_{\mathbf{q}'}^\dagger + \mathrm{e}^{\mathrm{i} (\mathbf{q}+\mathbf{q}') \cdot \mathbf{r}_j} a_{\mathbf{q}} a_{\mathbf{q}'} + \mathrm{H.c.} \right]\label{eq:fs_ham_int} \\
        H_1 &= \frac{1}{\sqrt{N}} \sum_{j=1}^N \sum_{\mathbf{q}} \hbar \xi_{\mathbf{q}} P_j^{(\mathrm{e})} \left(\mathrm{e}^{\mathrm{i} \mathbf{q} \cdot \mathbf{r}_j} a_{\mathbf{q}} +  \mathrm{H.c.}\right),
    \end{align}
    where the coupling constants are
    \begin{subequations}
        \begin{align}
            g_{\mathbf{q}} &= \frac{\nu_0}{2} \frac{q_\parallel}{q} \sqrt{\frac{\omega_0}{\omega_q}}, \\
            D_{\mathbf{q}\mathbf{q'}} &= \frac{g_{\mathbf{q}}g_{\mathbf{q'}}}{\omega_0},\\ \xi_{\mathbf{q}}&=\nu_\mathrm{e}\, \frac{q_\parallel}{q} \sqrt{\frac{1}{2\hbar m_\mathrm{e} \omega_q}} ,
        \end{align}
    \end{subequations}
   with $q_\parallel=(q_x^2+q_y^2)^{1/2}$.

   \subsection{Raman scattering rate}
    \begin{figure}[t]
        \centering
        \includegraphics[width=\textwidth]{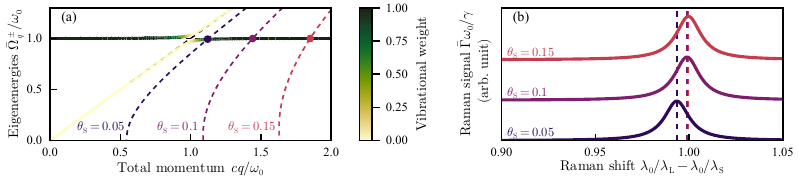}
        \includegraphics[width=\textwidth]{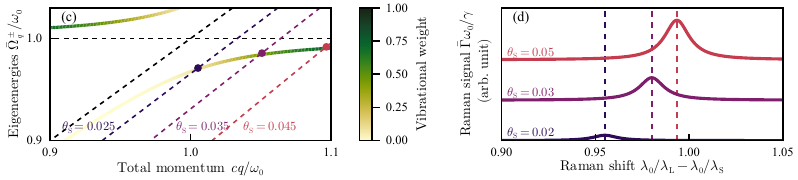}
        \caption{\label{fig:free_space_raman} 
        Raman spectra for molecules in free space. 
        Panels (a) and (c): Dispersion relation of the polaritonic modes from Eq.~\eqref{eq:Omega_free_space} (solid lines) and of the bare photon and vibrational modes (black dashed lines) for several scattered angles $\theta_\mathrm{S}$. 
        The color scale represents the vibrational weight $|x_{\mathbf{q}}^{\pm}|^2-|z_{\mathbf{q}}^{\pm}|^2$ of each polaritonic mode [cf.\ Eq.~\eqref{eq:weights_free_space}]. 
        Each colored dashed line corresponds to a specific $\theta_\mathrm{S}$ and is associated with the Raman spectra shown on the right. 
        Panels (b) and (d): Raman spectra for the corresponding angles $\theta_\mathrm{S}$ obtained by summing over both polaritonic branches. 
        Each peak is broadened by a Lorentzian of width $0.008\,\omega_0$.
        Parameters: $\omega_\mathrm{L} = 10.9\, \omega_0$, $\nu_0 = 0.066 \,\omega_0$ with $\lambda_0 = 2\pi c / \omega_0$ and analogous definitions for $\lambda_\mathrm{L}$ and $\lambda_\mathrm{S}$.}
    \end{figure}

    In order to diagonalize the Hamiltonian \eqref{eq:fs_ham_int}, we introduce the collective vibrational operators
    \begin{equation}
        S_{\mathbf{q}} = \frac{1}{\sqrt{N}} \sum_{j=1}^N \mathrm{e}^{-\mathrm{i} \mathbf{q} \cdot \mathbf{r}_j} b_j
    \end{equation}
    that satisfy the usual bosonic commutation relations, similarly to Sec.~\ref{subsec:bulk_cavity_diag}. Using these operators, we can rewrite Eq.~\eqref{eq:fs_ham_int} as $H^{(\mathrm{int})}_0 = \sum_{\mathbf{q}} H^{(\mathrm{int})}_{\mathbf{q}}$, where 
    \begin{equation}
        H^{(\mathrm{int})}_{\mathbf{q}} = \hbar \omega_q a_{\mathbf{q}}^\dagger a_{\mathbf{q}}^{\phantom{\dagger}} + \hbar \omega_0 S_{\mathbf{q}}^\dagger S_{\mathbf{q}}^{\phantom{\dagger}} - \mathrm{i} \hbar g_{\mathbf{q}}  \left(S_{\mathbf{q}}^\dagger a_{\mathbf{q}}^{\phantom{\dagger}} - S_{-\mathbf{q}} a_{\mathbf{q}} - \mathrm{H.c.}\right) + \hbar D_{\mathbf{q}\mathbf{q}} \left( a_{\mathbf{q}}^\dagger a_{\mathbf{q}}^{\phantom{\dagger}} - a_{\mathbf{q}} a_{-\mathbf{q}} + \mathrm{H.c.} \right).
    \end{equation}
    Introducing the Hopfield-Bogoliubov operators \cite{hopfield_theory_1958}
    \begin{equation}
        p_{\mathbf{q}}^{\pm} = w_{\mathbf{q}}^{\pm} a_{\mathbf{q}} + x_{\mathbf{q}}^{\pm} S_{\mathbf{q}} + y_{\mathbf{q}}^{\pm} a_{-\mathbf{q}}^\dagger + z_{\mathbf{q}}^{\pm} S_{-\mathbf{q}}^\dagger,
    \end{equation}
    we obtain two positive eigenvalues
    \begin{equation}
        \label{eq:Omega_free_space}
        \bar{\Omega}^{\pm}_{\mathbf{q}}=\omega_0\, \sqrt{\frac{1 + \chi_{\mathbf{q}} + {( \omega_{q}/\omega_0 )}^2}{2} \pm \sqrt{\frac{{( \omega_{q}/\omega_0)}^4 - 2 {( \omega_{q}/\omega_0 )}^2 \left(1 - \chi_{\mathbf{q}} \right) + \left(1 + \chi_{\mathbf{q}} \right)^2}{4}}},
    \end{equation}
    with the dimensionless coupling constant
    \begin{equation}
        \chi_{\mathbf{q}} = 4\left(\frac{g_{q}}{\omega_0}\right)^2 \frac{\omega_{q}}{ \omega_0} = \left(\frac{\nu_0}{\omega_0} \frac{q_\parallel}{q}\right)^2.
    \end{equation}
    The corresponding Hopfield-Bogoliubov coefficients are
    \begin{subequations}
        \label{eq:weights_free_space}
        \begin{align}
            w_{\mathbf{q}}^{\pm} &= \mathrm{i}\,\frac{\left(\omega_0 + \Omega^{\pm}_{\mathbf{q}}\right)\left(\omega_{q} + \Omega^{\pm}_{\mathbf{q}}\right)}{2 \omega_{q} g_{\mathbf{q}}} \,z_{\mathbf{q}}^{\pm},\\
            x_{\mathbf{q}}^{\pm} &= \frac{\omega_0 + \Omega^{\pm}_{\mathbf{q}}}{\omega_0 - \Omega^{\pm}_{\mathbf{q}}} \,z_{\mathbf{q}}^{\pm},\\
            y_{\mathbf{q}}^{\pm} &= \mathrm{i}\,\frac{\left(\omega_0 + \Omega^{\pm}_{\mathbf{q}}\right)\left(\omega_{q} - \Omega^{\pm}_{\mathbf{q}}\right)}{2 \omega_{q} g_{\mathbf{q}}} \,z_{\mathbf{q}}^{\pm},\\
            z_{\mathbf{q}}^{\pm} &= \sqrt{\frac{\omega_{q} g_{\mathbf{q}}^2 \left(\omega_0 - \Omega^{\pm}_{\mathbf{q}}\right)^2}{\Omega^{\pm}_{\mathbf{q}} {[\omega_0^2 - (\Omega^{\pm}_{\mathbf{q}})^2]}^2 + 4 \Omega^{\pm}_{\mathbf{q}} \omega_0 \omega_{q} g_{\mathbf{q}}^2}}.
        \end{align}
    \end{subequations}

    The Raman scattering transition element can be expressed by considering the initial state $\ket{\mathrm{I}} = a_{\mathbf{q}^\mathrm{L}}^\dagger \ket{\mathrm{G}}$ and the final state $\ket{\mathrm{F}} = p_{\mathbf{q}}^{\pm\dagger} a_{\mathbf{q}^\mathrm{S}}^\dagger \ket{\mathrm{G}}$, where $\ket{\mathrm{G}}$ is the ground state of the system. 
    Here, $\mathbf{q}^\mathrm{L}$ and $\mathbf{q}^\mathrm{S}$ are the laser and scattered photon wavevectors, respectively.
    Following the same procedure as in Sec.~\ref{subsec:raman_scattering}, we find that the total momentum is conserved, that is $\mathbf{q} = \mathbf{q}^\mathrm{L} - \mathbf{q}^\mathrm{S}$. The total Raman scattering rate reads
    \begin{equation}
        \bar{\Gamma} = \gamma \sum_{\sigma=\pm} \delta\!\left(\omega_\mathrm{L} - \omega_\mathrm{S} - \Omega^s_{|\mathbf{q}^\mathrm{L} - \mathbf{q}^\mathrm{S}|}\right) \left( x^s_{|\mathbf{q}^\mathrm{L} - \mathbf{q}^\mathrm{S}|} \right)^2,
    \end{equation}
    where $\gamma$ is defined similarly as in Sec.~\ref{subsec:raman_scattering}. 

    We plot in Figs.~\ref{fig:free_space_raman}(a) and \ref{fig:free_space_raman}(c) the dispersion relation from Eq.~\eqref{eq:Omega_free_space} over different ranges of the momentum $q$. The corresponding Raman spectra for different values of $\theta_\mathrm{S}$ are displayed in Figs.~\ref{fig:free_space_raman}(b) and \ref{fig:free_space_raman}(d). Such spectra are obtained by summing over the two possible final polaritonic states, that is over all possible scattered angles $\theta_\mathrm{S}$, and by broadening each peak by a Lorentzian of width $0.008\, \omega_0$ to reproduce experimental conditions. We observe only a single peak in the Raman spectra, which corresponds to the lower polariton branch, in qualitative agreement with the experimental results of Ref.~\cite{henry_raman_1965}.

\bibliography{biblio}